\documentclass[useAMS,usenatbib,twocolumn]{aastex631}

\usepackage[T1]{fontenc}
\usepackage{ae,aecompl}
 \usepackage{epsfig}
 \usepackage{amsmath}
 \usepackage{amssymb}
 \usepackage{natbib}
 \usepackage{multirow}
 \usepackage{graphicx}\graphicspath{{figures/}}
 \usepackage{color}
 \usepackage{epstopdf}
 \usepackage{float}
 \usepackage{tabularx}

\newcommand{\Msun}{\mathrm{M}_\odot}

\shorttitle{The Hateful Eight}
\shortauthors{Kimmig et al.}

\begin{document}
\title{The Hateful Eight: Connecting Massive Substructures in Galaxy Clusters like Abell~2744\\ to their Dynamical Assembly State using the Magneticum Simulations}

\correspondingauthor{Lucas C. Kimmig}
\email{lkimmig@usm.lmu.de}

\author{Lucas C. Kimmig}
\affil{Universit\"ats-Sternwarte M\"unchen, Fakult\"at f\"ur Physik, LMU M\"unchen, Scheinerstr.\ 1, D-81679 M\"unchen, Germany}

\author{Rhea-Silvia Remus}
\affil{Universit\"ats-Sternwarte M\"unchen, Fakult\"at f\"ur Physik, LMU M\"unchen, Scheinerstr.\ 1, D-81679 M\"unchen, Germany}

\author{Klaus Dolag}
\affil{Universit\"ats-Sternwarte M\"unchen, Fakult\"at f\"ur Physik, LMU M\"unchen, Scheinerstr.\ 1, D-81679 M\"unchen, Germany}
\affil{Max-Planck-Institute for Astrophysics, Karl-Schwarzschild-Str.\ 1, D-85748 Garching, Germany}

\author{Veronica Biffi}
\affil{Universit\"ats-Sternwarte M\"unchen, Fakult\"at f\"ur Physik, LMU M\"unchen, Scheinerstr.\ 1, D-81679 M\"unchen, Germany}
\affil{INAF - Osservatorio Astronomico di Trieste, via Tiepolo 11, SI-34143 Trieste, Italy}

\begin{abstract}
Substructures are known to be good tracers for the dynamical states and recent accretion histories of the most massive collapsed structures in the Universe, galaxy clusters. Observations find extremely massive substructures in some clusters, especially Abell~2744, which are potentially in tension with the $\Lambda$CDM paradigm since they are not found in simulations directly. However, the methods to measure substructure masses strongly differ between observations and simulations. Using the fully hydrodynamical cosmological simulation suite \textsc{Magneticum Pathfinder} we develop a method to measure substructure masses in projection from simulations, similar to the observational approach. We identify a simulated Abell~2744 counterpart that not only has eight substructures of similar mass fractions but also exhibits similar features in the hot gas component. This cluster formed only recently through a major merger together with at least 6 massive minor merger events since $z=1$, where prior the most massive component had a mass of less than $1\times10^{14}M_\odot$. We show that the mass fraction of all substructures and of the eighth substructure separately are excellent tracers for the dynamical state and assembly history for all galaxy cluster mass ranges, with high fractions indicating merger events within the last $2\,\mathrm{Gyr}$. Finally, we demonstrate that the differences between subhalo masses measured directly from simulations as bound and those measured in projection are due to methodology, with the latter generally 2-3 times larger than the former. We provide a predictor function to estimate projected substructure masses from \textsc{SubFind} masses for future comparison studies between simulations and observations.
\end{abstract}


\section{Introduction}
Galaxy clusters come in many different flavors: from really relaxed clusters with smooth hot X-ray halos and a clearly identifiable brightest cluster galaxy (BCG), like Abell~383 \citep[e.g.,][]{allen08}, to highly disturbed systems with detectable shock fronts and multiple massive galaxies, like Abell~2744 \citep[e.g.,][]{jauzac16} or MACS J0416.1-2403 \citep[e.g.,][]{grillo15}. Their dynamical states are commonly linked to their recent accretion history, and are thought to reflect the state of their cosmic environment given that they mark the nodes of the collapsing cosmic web.

Galaxy cluster substructures are some of the best indicators to provide insights into both the dynamical state \citep[e.g.,][]{delucia04,neto07,biffi16} and the accretion history \citep[e.g.,][]{jiang16} of their host galaxy clusters, as well as providing a test for potential dark matter variants \citep{bhatta21} and cosmological parameters \citep[e.g.,][]{ragagnin21}. 
Recent gravitational lensing obervations of galaxy cluster substructures have posed challenges to the $\Lambda$CDM paradigm, due to discrepancies found between the observations and cosmological simulations: The observed substructure masses are found to be larger \citep[e.g.,][]{jauzac16,schwinn17}, especially in the central regions \citep{grillo15}, and they appear more concentrated \citep[e.g.,][]{meneghetti20,ragagnin22}.
While the latter is not resolved purely from the inclusion of baryonic physics \citep{munari16}, it has been discussed to arise from the exact included baryonic subgrid physics in simulations \citep{bahe21}. The former, in turn, is thought to possibly arise from projection effects, albeit this has so far only been tested using dark matter only simulations \citep{mao18,schwinn18}.
Therefore, to enable future joint investigations employing both hydrodynamical simulations and gravitational lensing observations, it is necessary to understand the relationship between the intrinsic, three-dimensional substructure identification from simulations and the projected substructure masses, which we will analyze in this study.

An excellent testing ground for this endeavor is provided by the particularly extreme case of galaxy cluster Abell~2744 at $z=0.308$, with eight substructure masses measured by \citet{jauzac16} to all contain masses in excess of $5\times10^{13}\Msun$ within 150~kpc apertures, with the additional difficulty that those substructures are all located in close proximity within a sphere with a radius of $1\,\mathrm{Mpc}$.
This cluster has also been mapped in X-ray and radio bands, detecting both strong shock fronts \citep[e.g.,][]{owers11,eckert15} in combination with radio relics \citep[e.g.,][]{giovannini99,eckert16,rajpurohit21}. All these detection have been interpreted as the results of at least one massive recent merger event \citep[e.g.,][]{kempner04,boschin06}, if not even multiple merger events \citep[e.g.,][]{merten11}.
As part of the Hubble Frontier Field program \citep{lotz17}, Abell~2744 is one of the best studied and deepest imaged galaxy clusters, with multiple studies on its strong and weak lensing properties \citep[e.g.,][]{jauzac16,mahler18,bird18}. In addition, Abell~2744 is part of the GLASS-JWST program \citep{treu22} and as such more detailed studies on its properties are to be expected soon, especially with regard to strong lensing \citep{bergamini22}.

In this study, we aim to identify an Abell~2744 counterpart in a fully hydrodynamical cosmological simulation, with all baryonic physics included, and conclusively answer the question whether the large substructure masses observed by strong lensing measurements are really in tension with the $\Lambda$CDM paradigm or if this tension can be solved when accounting for projection effects. We further analyze whether projected substructure masses can still be used as a tracer for the accretion history of the host galaxy cluster.

To this end, we utilize the fully hydrodynamical cosmological simulation suite {\it Magneticum Pathfinder}, which is presented in Sec.~\ref{sec:sim}. In Sec.~\ref{sec:method} we introduce a method to identify substructures in projections similar to what is possible observationally, and compare the resulting substructure masses to what is obtained as bound subhalos directly from the simulation output in Sec.~\ref{sec:results}. The method will be used to identify Abell~2744 counterparts in the simulation in Sec.~\ref{sec:dynstat}, analyzing its formation pathways in Sec.~\ref{subsec:abell} and finally generalizing the results over all galaxy cluster mass ranges above $M_\mathrm{vir} \geq 1\times10^{14}\Msun$ in Sec.~\ref{subsec:accdyn}, connecting the projected substructure mass fractions to the dynamical state of galaxy clusters (Sec.~\ref{subsec:cshft}). Finally, we will summarize and conclude this study in Sec.~\ref{sec:summary}.

\section{Simulation}\label{sec:sim}
\sectionmark{Simulation}
The employed simulation is the fully hydrodynamical cosmological simulation {\it Magneticum Pathfinder}\footnote{www.magneticum.org} (Dolag et al., in prep.), following a WMAP-7 cosmology as $\Omega_0=0.272$ and $h=0.704$ from \citet{komatsu11}. 
All simulations were performed using an updated version of the Tree-PM SPH-code GADGET-2 \citep{springel05}, with SPH modifications according to \citet{dolag04,dolag05,donnert13,beck16}.
Employed physics include star formation, metal enrichment, and cooling processes \citep{tornatore04,tornatore07,wiersma09}, as well as AGN feedback by \citet{fabjan10,hirschmann14}. The details are discussed in more depth by \citet{teklu15}, and \citet{dolag17}. 
{\it Magneticum Pathfinder} reproduces global galaxy properties well, such as angular momentum \citep{teklu15,schulze18} and density distributions \citep{remus17_2,harris20,remus22}, properties of galaxies in cluster environments \citep{lotz19,lotz21}, as well as X-ray emission from galaxy clusters \citep{biffi18}.

The boxes cover a wide range of resolutions and sizes. As in this study the substructures of the most massive galaxy clusters are investigated, {\it Box2b/hr} is chosen because with a box volume of $(909\,\mathrm{cMpc})^{3}$ it is sufficiently large to contain galaxy clusters in excess of $M_\mathrm{FOF}>1\times10^{15}\Msun$ at a comparable redshift to Abell~2744. Simultaneously, with a mean stellar particle resolution of $4.97\times10^{7}\Msun$ it resolves halos with at least~100 stellar particles down to a total stellar mass of $M_\mathrm{*}\geq5\times10^{9}\Msun$. 
These galaxy clusters from the {\it Magneticum Pathfinder} simulation suit can also be accessed through the web portal (https://c2papcosmosim.uc.lrz.de), see \citet{ragagnin17} for more details.

At a redshift of $z \approx 0.252$, to be compatible with that of Abell~2744, halos were identified using the baryonic version of the halofinder \textsc{SubFind} \citep{dolag09} which uses a binding criterion to select particles belonging to the halos. To ensure a broad mass range of galaxy clusters considered, four mass bins of 29~galaxy clusters each are selected. The first, henceforth ``giants", comprises the most massive 29~galaxy clusters as defined by their friends-of-friends mass, where all have $M_\mathrm{FOF}>1\times10^{15}\Msun$. Then the bins ``medium", ``small" and ``tiny" are chosen such that their mean virial mass is tightly distributed around $5,2 \textnormal{ and } 1\times10^{14}\Msun$, respectively. The mass bins are summarized in Tab.~\ref{tab:mass_bins}.

\begin{table}[ht]
\caption{Summary of the mass bins used. Each contains the first 29~galaxy clusters within the specified mass range, as sorted by their friend-of-friends mass.}
\begin{center}
\begin{tabular}{ |c|c|c| }
\hline
     Mass Bin & Mass Range [$10^{14}\Msun$]  & $\bar{M}_\mathrm{vir}$ [$10^{14}\Msun$] \\
\hline
\rule{0pt}{0.9\normalbaselineskip} Giants & $M_\mathrm{FOF} > 10$& 13.13\\
     Medium & $ 5.15 > M_\mathrm{vir} > 4.85$    & 5.002\\
     Small  & $ 2.01 > M_\mathrm{vir} > 1.99$    & 2.000\\ 
     Tiny   & $ 1.002 > M_\mathrm{vir}>0.998$    & 1.000\\
\hline
\end{tabular}
\label{tab:mass_bins}
\end{center}
\end{table}

\section{Method: Cylinder Projection}\label{sec:method}
\sectionmark{Method}

The goal is to develop a method similar to the procedures from gravitational lensing observations, while remaining within the raw particle data and minimizing model assumptions. For such lensing analyses, one typically differentiates between a cluster-scale overdensity and many smaller galaxy-scale overdensities. These are commonly modeled by dPIE density profiles as introduced by \citet{kassiola93}. However, implementation of lenses via models within the simulations would require a rather large amount of additional assumptions. Instead, we rather assume that the lensing models can, as postulated, capture the real underlying mass distribution successfully and therefore focus more on the aspect of bound versus projected mass distributions: While from a simulation side it is rather simple to separate stellar and dark matter particles that form an individual bound (sub)structure from those particles that are only bound to the full structure potential, this is impossible in observations. This is especially hampered by the fact that observations only provide projected information by nature, representing one of the largest differences to simulations. Thus, we will in the following introduce the method used here to analyze the simulation as close to the observational procedure as possible without implementing model assumptions, and compare it to the information obtained directly from the simulation.\\[1em]

\subsection{Mass Maps}
The starting point for obtaining the projected substructure masses is a projected mass map. Mass particles from the simulation are directly projected onto a plane with a depth of $r_\mathrm{z}\approx 5\cdot r_\mathrm{vir}$ in front and behind the galaxy cluster. Even for the largest galaxy clusters of the ``giants'' this is more shallow than typical spectroscopic criterion employed in gravitational lensing analyses (see for example \citet{jauzac16} where the spectroscopic depth criterion equates to around $r_\mathrm{z}\approx38\,\mathrm{Mpc}\approx 15\cdot r_\mathrm{vir}$). 

As the projection angle is arbitrary, but the resulting quantities could vary strongly with projection, for each galaxy cluster~200 projections are sampled from an isotropical distribution on a spherical surface, resulting in 5800 galaxy clusters in projection in total per cluster mass bin. The initial projected map is centered on the most-bound-particle as determined by \textsc{SubFind}, which is a common choice of center for structure finders as this coincides with the deepest point of the potential well of the cluster. 

This position must {\it not} equal the center-of-mass, as that only coincides for a smooth, undisturbed, continuous mass distribution. Instead, the center-of-mass is determined from the projected total particle data directly via a shrinking sphere algorithm based on the work by \citet{power03}. The resulting barycenters are rather independent of initial parameters, so long as a sufficiently large initial area is used. Here the initial area is chosen as a circle of radius $1\cdot r_\mathrm{vir}$, and the barycenter is determined then in 2D. The shrinking factor is $f_\mathrm{shrink}=0.975$, as given by \citet{power03}, and a lower limit of~1000 particles within the circle is used as a break-off criterion for the algorithm. Alternatively, the algorithm ends if the barycenter varies over an iteration less than two times the softening length of the dark matter.

To highlight the differences, the first column of Fig.~\ref{fig:method} illustrates both the \textsc{SubFind} galaxy cluster center (black circle) and the projected center-of-mass (black cross) for two example galaxy clusters from the ``giants" mass bin, cluster~5 (top row) and~20 (bottom row), within an area of $(6\,\mathrm{Mpc})^2$. For the first cluster the projected center-of-mass is rather similar to the center from \textsc{SubFind}, while there exists a stark difference for the second cluster which is much more elongated and yet in a state of assembly. Here the deviation reaches on the order of half the virial radius. Therefore, this clearly highlights the importance of finding the center-of-mass when comparing to observations.\\[1em]

\subsection{Identifying Substructure}

Considering the substructure within the galaxy clusters, first a distinction must be made. The gravitationally bound structures as identified by \textsc{SubFind} are referred to in the following as {\it subhalos} (with masses written as $M_\mathrm{sub}$), while the projected masses within a given aperture are instead referred to as {\it substructure} (with masses written as $M_\mathrm{cyl}$). The latter may only appear close together, while being actually comprised of several separate structures. 

\begin{figure*}
  \begin{center}
  \includegraphics[width=0.98\textwidth]{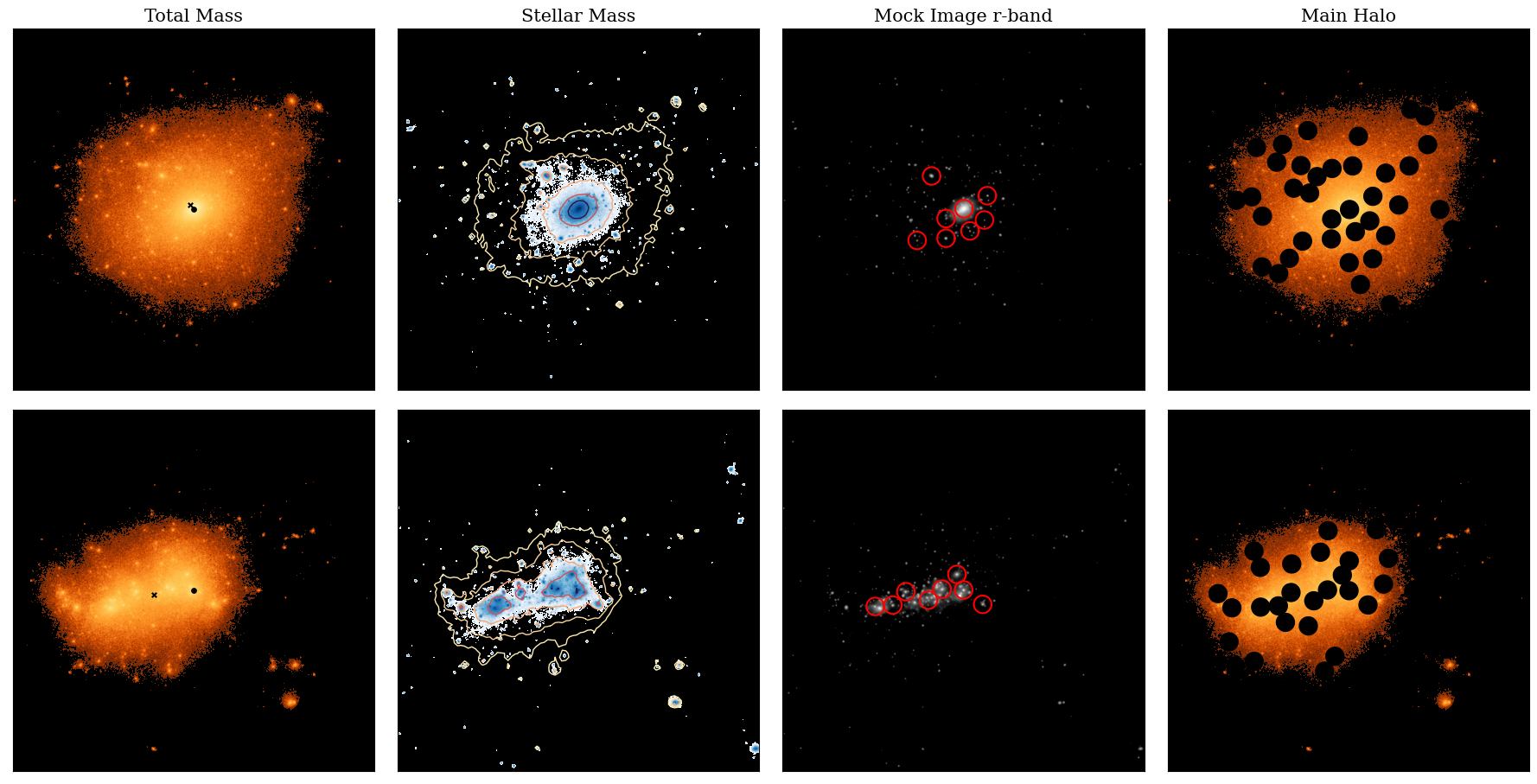}
   \caption{Examples of surface mass density and brightness maps for ``giant'' galaxy clusters~5 (\emph{top}) and~20 (\emph{bottom}). \emph{Left:} Projected total mass map, with the \textsc{SubFind} cluster center (i.e., the most-bound-particle position) indicated as a black dot and the projected center-of-mass marked with a cross. The colorbar range is $-3.7<\log_\mathrm{10}(\Sigma [10^{10}\Msun])<0.1$. \emph{Center left:} The stellar mass map, with isodensity lines of the total mass map. The colorbar range is $-6<\log_\mathrm{10}(\Sigma [10^{10}\Msun])<-1$. \emph{Center right:} Mock image in the r-band as seen from $z=0$ down to limiting magnitude of $25$~mag, generated via the method from \citet{martin22}. The red circles denote the eight most massive identified substructures. \emph{Right:} Total mass map with identified substructures masked out, leaving the main halo.}
  {\label{fig:method}}
  \end{center}
\end{figure*}

Potential positions of substructure are identified via the centers of subhalos from \textsc{SubFind}. To this end, all subhalos above a stellar mass cut $M_\mathrm{*,cut}$ are considered as initial positions. However, the subhalos may be elongated and as such in projection could appear to have a shifted center relative to their most-bound-particles. Similarly, two subhalos overlapping in projection may appear as a single structure with a center between them. As galaxy-scale models within gravitational lensing analyses are placed based on the observed maps, i.e., based on the stellar mass, this process is mimicked via use of a projected stellar mass map. The centers of substructures are then determined by converging to the most massive stellar pixel within a given radius $r_\mathrm{conv}$ around the subhalo centers. This ensures that at least one real bound structure is within the aperture, while simultaneously accounting for how the structures appear in projection. 

These potential substructure candidates are then sorted by their summed stellar masses within an area of $(9\cdot r_\mathrm{pix})^2=(18\,\mathrm{kpc})^2$ (scaled down for each bin as given in Tab.~\ref{tab:scaling}), and apertures are placed beginning with the most massive candidates. As apertures representing separate substructures should not overlap too strongly, this sorting ensures that the most massive obvious substructures are placed first. With an aperture radius of $r_\mathrm{ap}$, the choice of minimum distance to already placed apertures of $d_\mathrm{min}=\sqrt{3}\cdot r_\mathrm{ap}$ ensures that at no point can three apertures overlap simultaneously. 

\begin{table}[ht]
\caption{Overview of the relevant parameters for each mass bins used. Each mass bin is scaled relative to the ``giants" according to their relative mean virial mass. The stellar mass cut used throughout is $M_\mathrm{*,cut}$ as indicated on the left in Fig.~\ref{fig:subs}, and the projection depth is $r_\mathrm{z}\approx 5\cdot r_\mathrm{vir}$.}
\begin{center}
\begin{tabular}{ |r|cccc| }
\hline
     \rule{0pt}{0.9\normalbaselineskip} & Giants & \textnormal{Medium} & Small & Tiny \\
\hline
     \rule{0pt}{0.9\normalbaselineskip} $f_\mathrm{m,scale}$ & 1 & 0.381 & 0.152 & 0.076 \\
\hline
     \rule{0pt}{0.9\normalbaselineskip} $r_\mathrm{tot}$ [Mpc] & 1.3 & 0.942 & 0.694 & 0.551\\
     $r_\mathrm{ap}$  [kpc] & 150.0 & 109  & 80.1 & 63.6\\
     $r_\mathrm{conv}$[kpc] & 25.0  & 18.1 & 13.4 & 10.6\\
     $r_\mathrm{pix}$ [kpc] & 2.0   & 1.45 & 1.07 & 0.848\\
    \rule{0pt}{0.9\normalbaselineskip} $M_\mathrm{*,cut}$ [$10^{10}\Msun$] & 4.0 & 1.524 & 0.609 & 0.5\\
\hline
\end{tabular}
\label{tab:scaling}
\end{center}
\end{table}

Generally, the choice of aperture size $r_\mathrm{ap}$ is arbitrary, as one would need to know the mass of a substructure to obtain a representative radius for a structure of a given mass, while a radius is needed to measure the mass within to obtain a mass. Therefore, any given aperture could technically be chosen as a starting point. As one of the motivations to undertake this study is to test if we can reproduce the substructure properties of the galaxy cluster Abell~2744, we chose the same aperture size for the ``giants'' clusters as done by \citet{jauzac16}, namely $r_\mathrm{ap}=150\mathrm{kpc}$. As this would be unphysically large for the smaller clusters, we scale the size of the aperture accordingly to the mass ratio of the virial masses of the ``giants'' to the smaller bin, i.e.,
\begin{equation}
    r_\mathrm{ap} = 150\,\mathrm{kpc}\cdot\sqrt[3]{\bar{M}_\mathrm{vir,i} / \bar{M}_\mathrm{vir,giants}}=150\,\mathrm{kpc}\cdot\sqrt[3]{f_\mathrm{m,scale}}.
\end{equation}
The resulting values used for the different mass bins studied in this work can be found in Tab.~\ref{tab:mass_bins} and Tab.~\ref{tab:scaling}.

\begin{figure}
  \begin{center}
  \includegraphics[width=0.95\columnwidth]{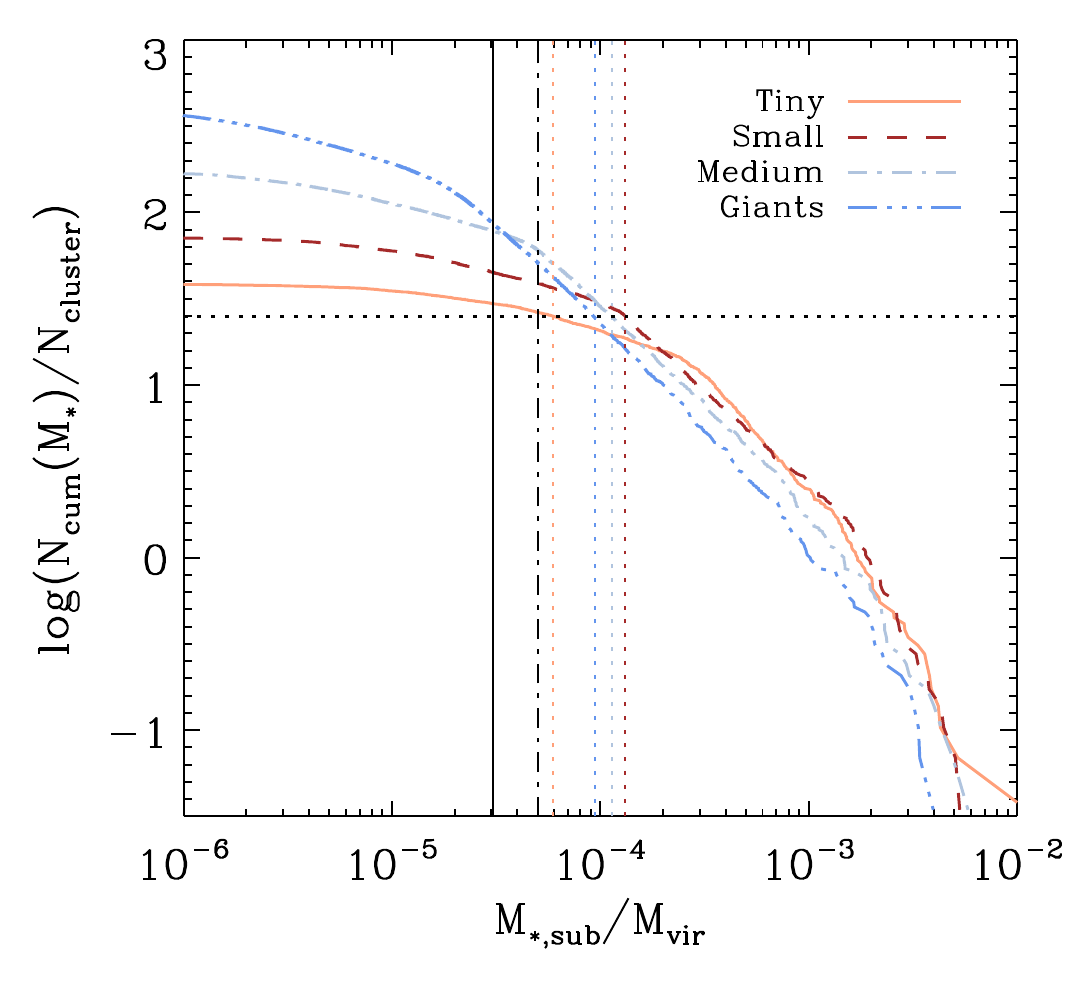}
  \caption{The logarithm with base 10 of the cumulative subhalo abundance per main halo in dependence of the subhalo stellar mass fraction of the main halo. Different colored lines represent the four mass bins. Vertical lines denote two different scalings of $M_\mathrm{*,cut}$. The solid black line scales linearly with the mean virial mass of the galaxy clusters, except for the ``tinies" (black dash dotted) as they reach the resolution level of at least~100 stellar particles. The colored vertical lines instead scale such as to ensure an approximately constant number of apertures ($\approx 25$, indicated by the horizontal line).
  }
  {\label{fig:subs}}
  \end{center}
\end{figure}

The stellar mass cut for potential candidates for substructures, $M_\mathrm{*,cut}$, is scaled linearly with the mean virial mass of the galaxy clusters, with a lower bound requiring that all substructures have sufficient particles given the simulation resolution limits. Fig.~\ref{fig:subs} shows the cumulative subhalo abundance for each of the four cluster mass bins, with the black solid line marking the stellar mass cut. Only for the smallest mass bin, the ``tiny'' clusters, this threshold needs to be shifted slightly to preserve the limit of at least 100 particles per subhalo, marked by the dash-dotted black line. While on the one hand it can be seen that the subhalo number distributions of the clusters are generally self-similar, it can also be seen that the more massive galaxy clusters have significantly more subhalos with this threshold scheme, in agreement with actual observed galaxy clusters. For comparison the horizontal dotted line marks a constant~25 subhalos, with the resulting mass ratio for each cluster mass bin marked by the vertical lines in the respective colors of the cluster mass bins.

Both middle columns of Fig.~\ref{fig:method} demonstrate the process of finding substructures, again for cluster~5 (top row) and~20 (bottom row) of the ``giants": The second column shows the total stellar mass map, which in structure is very similar to the total mass maps shown on the left. Substructures are first identified as the centers of the subhalos above the stellar mass cut $M_\mathrm{*,cut}$ as per Tab.~\ref{tab:scaling}, and then shifted to the most massive pixels within the projected stellar mass map in the vicinity. The third column then depicts the resulting eight most massive substructures via red circles in a mock image representing what could be observed in the r-band with a magnitude down to $26$~mag using the method from \citet{martin22}. All eight substructures include the brightest galaxies of both clusters visible in the r-band. However, there are significantly more substructures identified overall, as can be seen in the right panel of Fig.~\ref{fig:method}, where all substructures are marked by the black filled circles.

\subsection{The Main Halo}
Finally, the contribution to the total mass from the main halo needs to be determined for every projection. This is done by first masking the substructures, as demonstrated in the right column of Fig.~\ref{fig:method}. Subsequently, the remaining particles are binned in 2D equal-mass bins and the density within concentric rings is calculated, while subtracting the mean density to include only overdensities. The resulting profile represents the main halo in projection and is found to be best fit by an Einasto profile \citep{einasto65}, which is fit in the form:
\begin{equation}
    \rho(r) = \rho_{-2}~\exp \left\{ -\frac{2}{\alpha_\mathrm{Ein}}~\left[~\left(\frac{r}{r_{-2}}\right)^{\alpha_\mathrm{Ein}} -1\right]~\right\},\label{eq:einasto}
\end{equation}
as given by \citet{retana12}. The added flexibility from the slope parameter $\alpha_\mathrm{Ein}$ allows better reproduction of the broad range of density profiles than the NFW profile \citep{navarro96}, even compared to fitting its projected form as given by \citet{takada03}.

Finally, the contribution from the mean background density $\rho_\mathrm{mean}$, which is given by the sum of all mass divided by the box volume, must be subtracted as well. Consequently, what remains within the apertures can be attributed solely to the substructures. The mass of substructures is then given by 
\begin{equation}
    \begin{aligned}
    M_\mathrm{cyl,i} &\equiv M^\mathrm{substructure}_\mathrm{ap,i} \\
    &= M_\mathrm{ap,i}- [\rho_\mathrm{Ein,fit}(d_\mathrm{i})+\rho_\mathrm{mean}] \cdot \pi r_\mathrm{ap}^2 \cdot r_\mathrm{z},
    \end{aligned}
    \label{eq:cylmass}
\end{equation}
with $M_\mathrm{ap,i}$ the summed mass of all particles within aperture $i$, $\rho_\mathrm{Ein,fit}(d_\mathrm{i})$ the main halo density at the center of the aperture and $\pi r_\mathrm{ap}^2 \cdot r_\mathrm{z}$ the volume of the projected cylinder. 

\section{Bound Versus Projected Masses}\label{sec:results}
\sectionmark{Results}

As mentioned, one of the major obstacles when comparing simulations with observations is the different dimensionality of the studied objects. Using the method outlined in Sec.~\ref{sec:method}, we can directly compare the values determined from the projection method with those resulting from bound structure via \textsc{SubFind}, thereby providing an estimate on how accurately the real underlying bound structures can be reconstructed from the~2D information.

\subsection{Total Mass in Projection Versus Bound Mass}\label{subsec:subfind}

\begin{figure}
  \begin{center}
  \includegraphics[width=0.95\columnwidth]{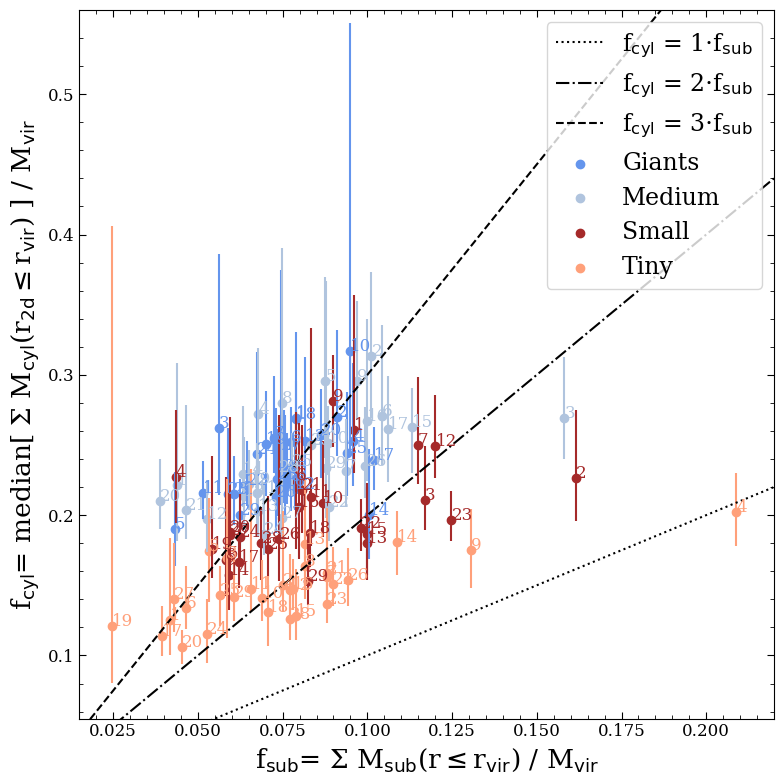}
  \caption{The median total substructure mass fraction $f_\mathrm{cyl}$ as a function of the total subhalo mass fraction $f_\mathrm{sub}$, colored by the mass bin. Colored lines denote the $1\sigma$ range of $f_\mathrm{cyl}$. The black lines denote a factor of 1 (dotted), 2 (dash-dotted) and 3 (dashed) between $f_\mathrm{cyl}$ and $f_\mathrm{sub}$.}
  {\label{fig:mcylmvir}}
  \end{center}
\end{figure}

The total mass fraction contained in substructures, i.e., in projected cylinder apertures, within a given radius relative to the total mass of the galaxy clusters is given as:
\begin{equation}
    f_\mathrm{cyl}(r)\equiv\left(\sum_\mathrm{i=2}^N M_\mathrm{cyl,i}(r_\mathrm{i}\leq r)\right)/M_\mathrm{tot}(\leq r)
\end{equation}
where the most massive substructure is defined as the main and thus excluded. This fraction is projection dependent. Note that we primarily care for the value at the virial radius and thus define $f_\mathrm{cyl}$ when given without an explicit radius as the value at $r_\mathrm{vir}$. Accordingly, the mass fraction contained within subhalos $f_\mathrm{sub}$ is calculated as the sum of all subhalos within the virial radius excluding the most massive one, divided by the virial mass. This fraction is based on the full three-dimensional information and as such is a fixed quantity independent of projection. A comparison between both fractions is shown in Fig.~\ref{fig:mcylmvir}, with median values of $f_\mathrm{cyl}$ as a function of $f_\mathrm{sub}$, colored by mass bin of the galaxy cluster. 

Generally, we find the projected masses to be larger by a factor around 2 to 3 than those determined from what is physically bound in local structures in three-dimensions. The scatter for individual galaxy clusters in $f_\mathrm{cyl}$ can be fairly large, as visible from the $1\sigma$ ranges.
Overall, the more massive clusters like ``giants'' and ``medium'' tend to have larger substructure mass fractions at fixed subhalo mass fraction lying around $20$-$30\%$. They thus overestimate the amount of mass within self-bound substructures by a factor of~3 on average. While this could simply be due to the fact that there are more substructures in those massive clusters than in the smaller ones, it also cautions the interpretation of such signals from observations as the same is not true for the $f_\mathrm{sub}$ values.

\begin{figure}
  \begin{center}
  \includegraphics[width=0.975\columnwidth]{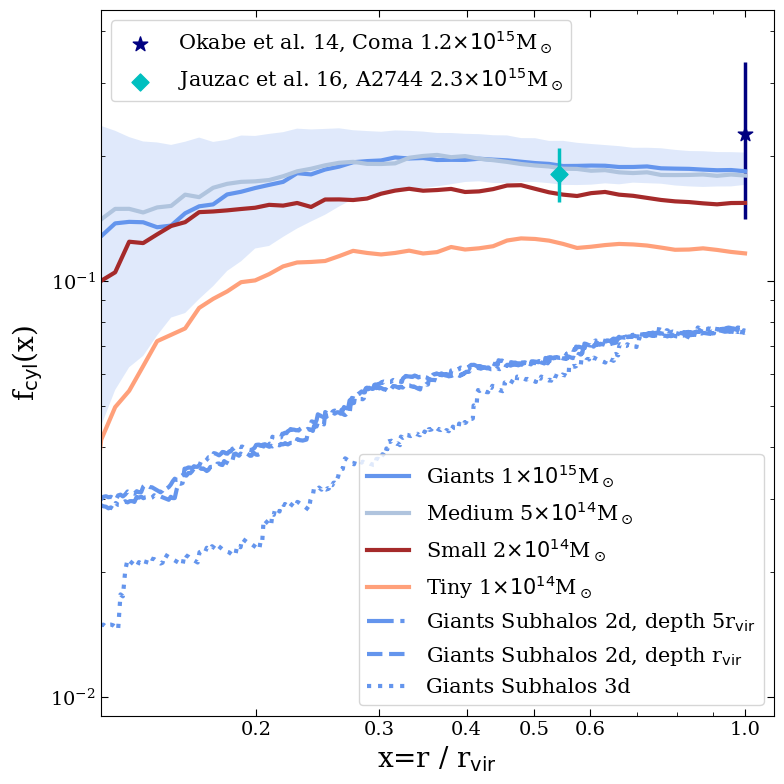}
  \caption{The radial dependence of the total mass within substructures as a fraction of total mass. For each mass bin (colored lines) the mean over all galaxy clusters is shown, where for the ``giants'' additionally the $1-\sigma$ range is plotted (light blue area). Also for the ``giants'' is shown the radial behavior of the bound subhalo mass fraction $f_\mathrm{sub}$, once in 3d (blue dotted) and once when projecting out to $1\cdot r_\mathrm{vir}$ (blue dashed) and $r_\mathrm{z}$ (blue dash-dotted). The cyan star and purple diamond represent values from lensing observations of the Abell~2744 and Coma Cluster, respectively, with vertical lines denoting the errors. Note that for the 3d curve $x$ is the fractional 3d distance to the cluster center, while for all others it is the fractional projected distance.}
  {\label{fig:radial}}
  \end{center}
\end{figure}

Another important characterization of galaxy clusters is the radial distribution of substructure masses. Fig.~\ref{fig:radial} depicts the radial behavior of the substructure mass fraction $f_\mathrm{cyl}$, where within a given radius $r$ the summed substructure mass is divided by the total mass contained within this radius. The total mass here is comprised of the substructure mass plus the integrated main halo mass as given by the Einasto fit from Eq.~\ref{eq:einasto}. 
At all radii and all cluster mass bins, we find that $f_\mathrm{cyl}$ is nearly constant for all four cluster mass bins, with fractions only dropping below $0.2\cdot r_\mathrm{vir}$. Between galaxy cluster mass bins, $f_\mathrm{cyl}$ increases with the mass of the galaxy cluster from the ``tiny'' to ``medium'' mass bin, but interestingly does not increase further to the most massive clusters and instead converges slightly below $f_\mathrm{cyl}=20\%$. 

These projected values are different to the values obtained when only considering the bound mass, as shown for the clusters of the ``giants'' mass bin in Fig.~\ref{fig:radial} as an example: the median radial behavior of the subhalo mass fraction $f_\mathrm{sub}$ is shown (non-solid blue lines), where the bound subhalos from \textsc{SubFind} are radially summed up and divided by the total mass within the radius. The dotted line shows this in 3d and it lies generally the lowest, in particular toward the center as there are very few bound subhalos at $r_\mathrm{3d}<0.2\cdot r_\mathrm{vir}$. In projection out to $1\cdot r_\mathrm{vir}$ in front and behind the cluster center (blue dashed line), however, it is possible to have bound subhalos which simply {\it appear} close to the center but actually lie farther out. Correspondingly, the curve is much flatter towards the center as here $x=r_\mathrm{2d}/r_\mathrm{vir}$ is instead the projected distance. Nonetheless, the values close to the virial radius converge to the 3d values as the few additional subhalos within the corners of the projected cylinder compared to the sphere make no discernible difference. The radial subhalo mass fractions found for this kind of projection are comparable with those predicted by \citet{jiang17} from a dark matter only simulation set.

When projecting out to $r_\mathrm{z}\approx5\cdot r_\mathrm{vir}$ (blue dash-dotted line) the subhalo mass fraction is nearly indistinguishable from the projection out to $1\cdot r_\mathrm{vir}$. The total mass fraction contained in bound subhalos at~$10\%$ thus still lies noticeably below the values we find when projecting the full particle distribution by around a factor of~2. Consequently, projection of bound subhalos alone is insufficient to explain the high substructure masses, and instead portions of their mass are the result of contributions from the main halo.

For comparison, the substructure mass fractions measured from lensing for the Coma cluster \citep[][purple star]{okabe14} and  Abell~2744 \citep[][cyan diamond]{jauzac16} are included in Fig.~\ref{fig:radial}. With respect to their mass, both are comparable to the ``giants'' mass bin (dark blue curve). Within the error bars we find excellent agreement between our prediction and the observations, which is surprising given that Abell~2744 has an extremely high fraction of mass in substructures in comparison to other galaxy clusters. However, this could be due to the fact that only the amount of mass in the eight most massive substructures is considered here, neglecting all other smaller substructures. Adding more measurements in the future will enhance our understanding of typical substructure distributions in galaxy clusters.

\subsection{Individual Substructure Masses}
While on average the projection increases the total projected substructure mass relative to the total bound subhalo mass by around a factor of~2, the question arises how strongly this scatters for individual measured substructures. In the following, we compare the individual projected masses to those of the bound subhalos within the same aperture. The majority of apertures contain just a single subhalo, though some can contain in excess of five. Defining the bound subhalo masses as $M_\mathrm{sub,j}$, then
\begin{equation}
    M^\mathrm{i}_\mathrm{sum\ sub\ in\ cyl} \equiv \sum M_\mathrm{sub,j\ in\ i}
\end{equation}
is the sum of all subhalos within an aperture $i$. In most cases even when multiple subhalos are present the total bound subhalo mass within an aperture is dominated by a single subhalo, where
\begin{equation}
    M_\mathrm{s} \equiv \mathrm{max}\{M_\mathrm{sub,j\ in\ i}\}
\end{equation}
is the mass of this most massive subhalo within the aperture. 
\begin{figure*}
  \begin{center}
  \includegraphics[width=0.98\textwidth]{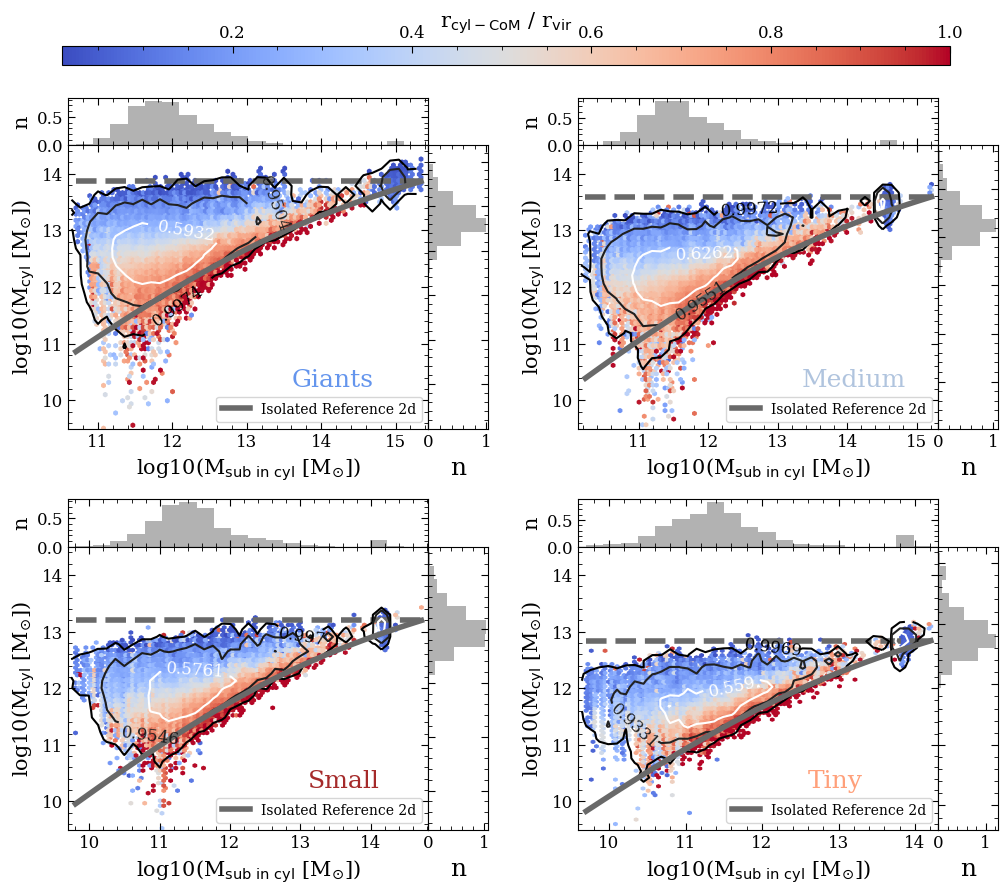}
  \caption{The substructure mass as a function of the associated subhalo mass for the four mass bins (as written), colored by the projected distance of the aperture to the center-of-mass. 
  The solid (dashed) gray lines indicate the according cylindrical aperture mass for the equivalent (maximum) halo mass in isolation. See text for details.
  The contour lines contain the fraction of total substructures as written. Histograms depict the number density $n$ distribution of either $M_\mathrm{s}$ (\emph{top}) or $M_\mathrm{cyl}$ (\emph{right}).}
  {\label{fig:msmc}}
  \end{center}
\end{figure*}

Thus, Fig.~\ref{fig:msmc} shows the substructure mass $M_\mathrm{cyl}$ as a function of $M_\mathrm{s}$ for the four cluster mass bins in separate panels. All galaxy cluster mass bins exhibit very similar distributions, with the projected mass on average increasing with increasing subhalo masses. The scatter is much larger for subhalos of smaller masses, with the projected aperture masses spreading over multiple magnitudes for the lowest mass subhalos. While we find that generally the mean substructure masses increase with the contained subhalo mass, the most massive substructure masses exhibit barely any dependence on the mass within the aperture which is bound in individual subhalos, instead only very weakly increasing with $M_\mathrm{s}$.

Assuming that a significant portion of this scatter in the $M_\mathrm{cyl}-M_\mathrm{s}$ correlation is caused by falsely adding contributions from the main halo, we should find a trend that those apertures closer to the center-of-mass experience stronger increases in $M_\mathrm{cyl}$. This can indeed be seen in Fig.~\ref{fig:msmc} where the color encodes the projected distance of the center of the aperture to the center-of-mass normalized by the virial radius, $r_\mathrm{2d}\equiv (r_\mathrm{cyl-CoM})/r_\mathrm{vir}$. Apertures which are very close to the center-of-mass (colored blue) lie at consistently high substructure masses. Conversely, substructures far out (red) are very strongly dependent on $M_\mathrm{s}$, following instead more the curvature of the solid gray curve. 

This solid gray curve represents the mass within an aperture $r_\mathrm{ap}$ which would result from a single isolated halo that follows an NFW-profile -- with mass $M_\mathrm{s}$ and concentration following the concentration-mass relation from \citet{ragagnin19} -- when projected out to a depth $r_\mathrm{z}$. This curve provides a lower bound to $95\%$ of the substructures for all cluster mass bins, as can be seen from the contours. An upper bound on the other hand is given by the dashed gray line. This is determined in the same way as the solid curve but for the most massive $M_\mathrm{s}$ within the cluster mass bin, and thus represents the substructure mass that one would get from projecting the center of the main halo, i.e., the highest-density area within the galaxy clusters. 

\begin{figure*}
  \begin{center}
  \includegraphics[width=0.98\textwidth]{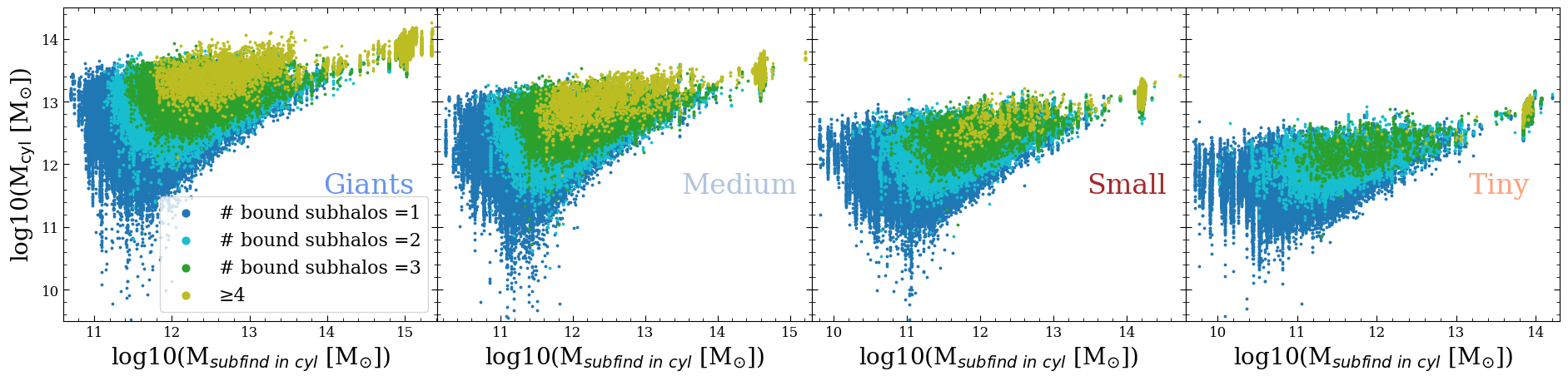}
   \caption{The substructure mass as a function of the most massive bound subhalo within its aperture as given by \textsc{SubFind} for the four mass bins (indicated within the plots), with colors denoting the number of bound subhalos within the substructures. Note that the apertures containing a higher number of subhalos are plotted over those with less to better show their behavior.}
  {\label{fig:msmc_snum}}
  \end{center}
\end{figure*}

The fact that the main bulk of substructure masses lie on or above the solid gray curve implies that the projected mass is in general \emph{not} carried by a single subhalo. Instead, contributions from in particular the main halo as well as from other subhalos are relevant. The latter can be seen in Fig.~\ref{fig:msmc_snum} where the substructures which contain higher numbers of subhalos are colored to the front. Those which contain these higher numbers are found in the top right of the $M_\mathrm{cyl}-M_\mathrm{s}$ plane, so for substructures which are more massive. This, however, does not imply that all massive substructure must contain multiple subhalos, as there are apertures containing just one subhalo within the top right which are simply overplotted. Instead, the dependency on projected distance seen in Fig.~\ref{fig:msmc} and thus the main halo contribution is more significant to the final substructure mass, with the number of included subhalos being a secondary effect.

\subsection{Quantifying $M_\mathrm{cyl}-M_\mathrm{s}$}

Based on the strong radial dependence as $M_\mathrm{cyl}=M_\mathrm{cyl}(M_\mathrm{s},r_\mathrm{2d})$ it is possible to construct a predictor function for the range of projected substructure masses that can arise for a given bound mass at a given projected distance to the cluster center. The parameter space is split into $n\times m$ bins in  $M_\mathrm{s}$ and $r_\mathrm{2d}$. Within each bin the distribution of substructure masses $M_\mathrm{cyl}$ follows a Gaussian (see Fig.~\ref{fig:mixture} in the Appendix for an example):
\begin{equation}
    f(x=M_\mathrm{sub};\mu,\sigma) = \frac{1}{\sqrt{2\pi\sigma^2}}\cdot\exp\left\{-\frac{(x-\mu)^2}{2\sigma^2}\right\},
\end{equation}
indicating that other possible systematics are comparatively minor (as the Gaussian represents a random distribution). This holds true so long as the number of bins is sufficiently high, $(n,m)\geq10$, as a low number of bins smears many different substructures over each other resulting in a distribution which is instead comprised of a sum of Gaussians. Here $(n,m)=(50,50)$ is chosen throughout. 

The fit parameters of the Gaussian are then the mean $\mu(M_\mathrm{s},r_\mathrm{2d})$ and variance $\sigma(M_\mathrm{s},r_\mathrm{2d})$. We find that the variance does not depend strongly on the projected distance, and is fit best by the form
\begin{equation}
    \sigma(M_\mathrm{s}) = \gamma\cdot \log_\mathrm{10}(M_\mathrm{s}[\Msun])+\sigma_\mathrm{0}.\label{eq:sigma}
\end{equation}
This can be seen in Fig.~\ref{fig:msmc} as the scatter is mainly dependent on $M_\mathrm{s}$, with substructures containing low mass subhalos scattering strongly in $M_\mathrm{cyl}$ while those with higher $M_\mathrm{s}$ scatter less, irregardless of the projected distance. Consequently, $\gamma<0$.

As for the means $\mu$, one can consider a bin in $r_\mathrm{2d}$, for example $r_\mathrm{2d}<0.2$. Substructures located at these distances will primarily occupy the dark blue regions in Fig.~\ref{fig:msmc}. Splitting this region into $M_\mathrm{s}$ bins (so vertical slices in Fig.~\ref{fig:msmc}) and determining the means $\mu$ finds them weakly linearly increasing in logspace with $M_\mathrm{s}$, so
\begin{equation}
    \log_\mathrm{10}(\mu)=\alpha\cdot \log_\mathrm{10}(M_\mathrm{s}[M_\odot])+\beta,\label{eq:mu_base}
\end{equation}
with $\alpha,\beta>0$. If considering the same for a bin of larger $r_\mathrm{2d}$, for example the dark red region, then the slope becomes larger and the y-intercept decreases. This dependency of $a,b$ on the projected distance can be fit well by a line, such that
\begin{align}
    \alpha(r_\mathrm{2d})&=\alpha_\mathrm{r}\cdot r_\mathrm{2d}+\alpha_\mathrm{0},\label{eq:alpha}\\
    \beta(r_\mathrm{2d})&=\beta_\mathrm{r}\cdot r_\mathrm{2d}+\beta_\mathrm{0},\label{eq:beta}
\end{align}
with $\alpha_\mathrm{r}>0$ and $\beta_\mathrm{r}<0$. Note that $\alpha_\mathrm{0}$ describes the distance-independent relationship between $\mu-M_\mathrm{s}$ and is thus expected to be positive, while $\beta_\mathrm{0}$ is the overall y-intercept of the line fits and thus should be on the order of typical values of $\mu$.
\begin{figure*}
  \begin{center}
  \includegraphics[width=0.98\textwidth]{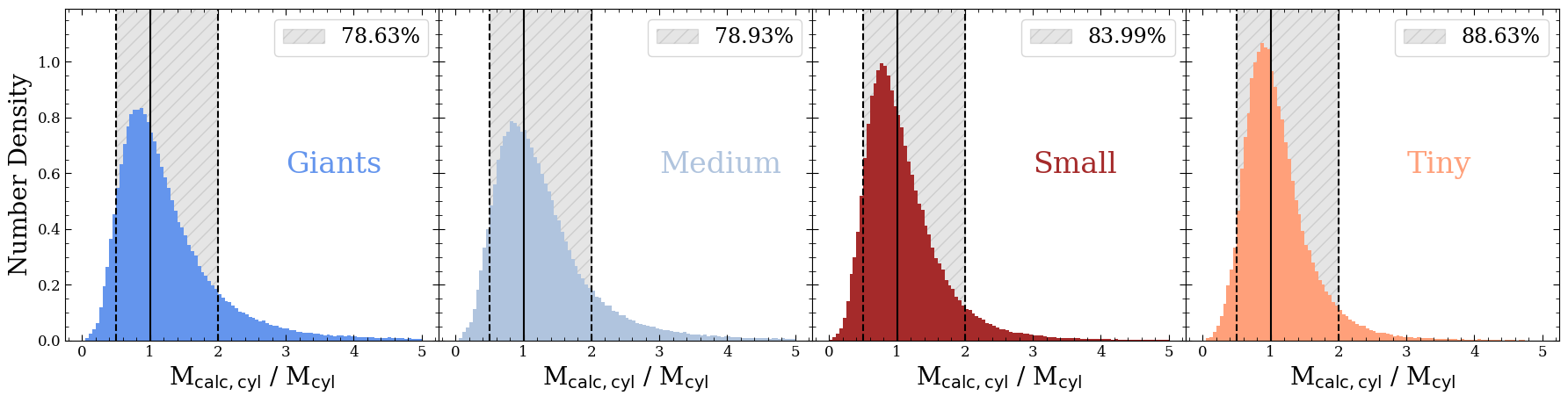}
  \caption{The distribution of ratios between predicted $M_\mathrm{cal,cyl}$ to measured substructure mass $M_\mathrm{cyl}$, for each of the four mass bins (indicated within the plots). The prediction is made via Eq.~\ref{eq:fit} and the parameters from Tab.~\ref{tab:para}. Solid vertical black line denotes equality, with the gray shaded area enclosing those within factor~2, with the number of substructures predicted within this range given in the legends.}
  {\label{fig:reproduction}}
  \end{center}
\end{figure*}

Combining Eq.~\ref{eq:mu_base} with Eq.~\ref{eq:alpha} and~\ref{eq:beta} gives the mean substructure mass as a function of projected radius and bound mass as:
\begin{equation}
    \begin{aligned}
    \log_\mathrm{10}(\mu(M_\mathrm{s},r_\mathrm{2d}))= &(\alpha_\mathrm{r}\cdot r_\mathrm{2d}+\alpha_\mathrm{0})\cdot\log_\mathrm{10}(M_\mathrm{s}[M_\odot])\\ 
    &+ \beta_\mathrm{r}\cdot r_\mathrm{2d}+\beta_\mathrm{0}.\label{eq:fit}
    \end{aligned}
\end{equation}

The resulting parameters for the Gaussians are then two for the variance $(\gamma,\sigma_\mathrm{0})$, and four for the means $(\alpha_\mathrm{r},\alpha_\mathrm{0},\beta_\mathrm{r},\beta_\mathrm{0})$. They are summarized for the different mass bins in Tab.~\ref{tab:para}.

\begin{table}
\caption{Summary of the predictor function parameters for the four mass bins.}
\begin{center}
\begin{tabular}{ |c|cc|cccc| }
\hline
     Mass Bin & $\gamma$ & $\sigma_\mathrm{0}$ & $\alpha_\mathrm{r}$ & $\alpha_\mathrm{0}$ & $\beta_\mathrm{r}$ & $\beta_\mathrm{0}$ \\
\hline
\rule{0pt}{0.9\normalbaselineskip} Giants & -0.0683 & 1.05 & 0.231 & 0.206 & -3.76 & 10.8\\
     Medium & -0.0666 & 1.01 & 0.218 & 0.226 & -3.42 & 10.2\\
     Small  & -0.0557 & 0.834 & 0.250 & 0.216 & -3.72 & 9.96\\ 
     Tiny   & -0.0548 & 0.789 & 0.323 & 0.210 & -4.43 & 9.80\\
\hline
\end{tabular}
\label{tab:para}
\end{center}
\end{table}

Using these parameters allows to predict the distribution of possible projected substructure masses which result from some bound subhalo of mass $M_\mathrm{s}$ located at projected distance $r_\mathrm{2d}$ from the projected center-of-mass. Note that these parameters vary with different masses of the host galaxy cluster (i.e., between cluster mass bins). The ratio of predicted mean mass of each substructure, $\mu\equiv M_\mathrm{calc,cyl}$, to actual measured mass $M_\mathrm{cyl}$ is plotted in Fig.~\ref{fig:reproduction}. Around $80\%$ of substructure masses are predicted within factor~2 of their real values, as highlighted in gray.

In summary, we find that on average the projected substructure masses within apertures of a given size are about a factor of 2-3 larger than the masses that are actually bound in subhalos, for all cluster mass ranges. Furthermore, the radial distance to the cluster center plays a crucial role in the amount of additional mass added to the individual substructures, in that substructures closer to the center may contain matter from the main halos that did not get perfectly subtracted. Moreover, some apertures can also contain more than one subhalo, adding to the overestimate of the mass. We provide a predictor function for future comparison studies between simulations and observations for the mass contained inside a substructure calculated from the subhalo mass and its distance from the main halo center that allows to predict the expected overestimate of the mass in projection without actually having to go through the process of projecting the cluster from simulations.
We conclude that it is necessary to take into consideration projection effects but especially the fact that observationally it is not possible to separate bound from unbound material when comparing simulations and observations, to ensure a fair comparison.

\section{Substructures and Cluster Dynamical State}\label{sec:dynstat}
\sectionmark{Abell 2744}
The existence of multiple massive substructures inside a galaxy cluster is thought to be connected to recent merging activity, as for example discussed for the case of the galaxy cluster Abell~2744 which is especially well known for its multiple massive substructures \citep{jauzac16}. Therefore, we will test this hypothesis in the following by first reproducing Abell~2744 from the simulations, using the method described in Sec.~\ref{sec:method} accounting for the projection effect, and then tracking its formation pathways. Furthermore, we then broaden the analysis to our full cluster sample, aiming to encode the information hidden in these substructures.

\begin{figure*}
  \begin{center}
  \includegraphics[width=0.975\textwidth]{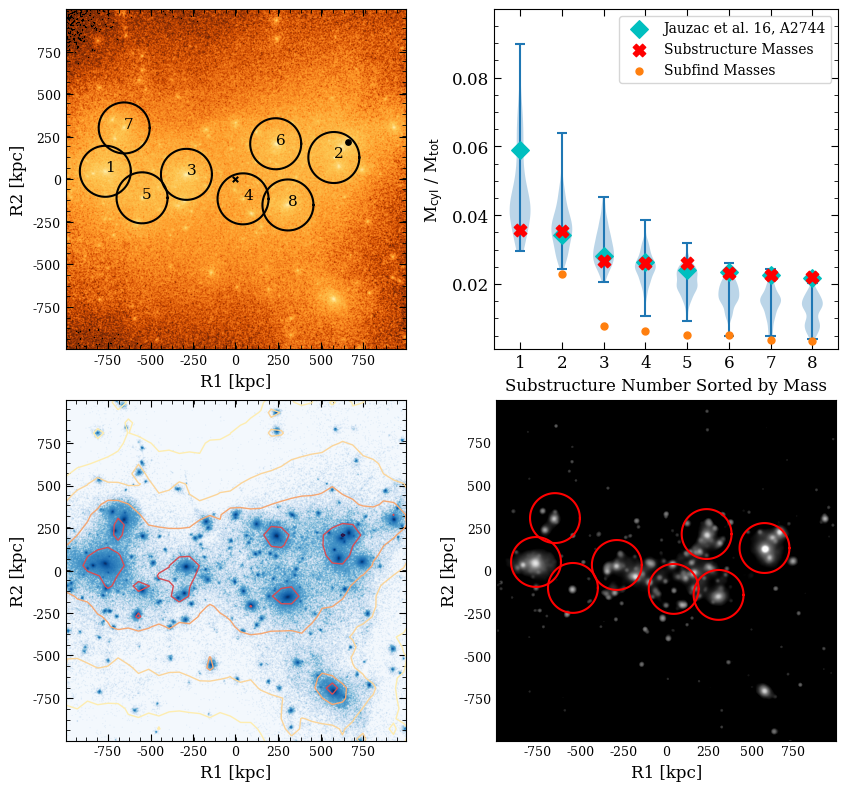}
  \caption{\emph{Top left:} Surface mass density map of the central $(2\,\mathrm{Mpc})^2$ region of galaxy cluster~20 of the ``giants'' for the best projection found overall, centered on the projected center-of-mass (black cross) with the \textsc{SubFind} cluster center indicated as a black dot. The eight most massive substructures are shown as black circles numbered decreasing with mass. \emph{Top right:} The substructure masses as a fraction of the total mass for the best projection (red crosses) as compared to those in Abell~2744 (blue diamonds) determined by \citet{jauzac16}. Underlayed in blue is the total distribution of masses for all~200 random projections of the galaxy cluster, with the width of the shaded area denoting their relative frequency. The subhalo masses from \textsc{SubFind} (orange dots) are given as a fraction of the $M_\mathrm{tot}$ for the best projection. \emph{Bottom left:} Same as \emph{top left} but for the stars, with isodensity contours from the total mass map overlayed. \emph{Bottom right:} Mock image in the r-band as seen from $z=0$ down to limiting magnitude of $25$~mag, generated via the method from \citet{martin22}. The red circles denote the eight most massive identified substructures. Colorbar ranges for the surface density maps are the same as in Fig.~\ref{fig:method}.}
  {\label{fig:Cl20_A}}
  \end{center}
\end{figure*}

\subsection{The case of Abell 2744}\label{subsec:abell}
Using gravitational lensing measurements, \citet{jauzac16} found extremely large masses inside substructures for the galaxy cluster Abell~2744, significantly more than can be found within simulations \citep{schwinn17}. Using dark-matter only simulations, \citet{mao18} and \citet{schwinn18}  present evidence that the discrepancy is caused by projection effects. Here, we will use, for the first time, a fully baryonic hydrodynamic cosmological simulation to identify Abell~2744 counterparts, applying the method outlined in Sec.~\ref{sec:method}. 

The eight substructures found by \citet{jauzac16} within Abell~2744 are not only very massive but also in close proximity within $1\,\mathrm{Mpc}$. They find that even the eighth most massive substructure still has a mass of $M_\mathrm{cyl,8}=5\times10^{13}\Msun$ within an aperture of $r_\mathrm{ap}=150\,\mathrm{kpc}$ (see Tab.~\ref{tab:frac_mass}). When compared to their measurement of the total mass within a radius $r_\mathrm{tot}=1.3\,\mathrm{Mpc}$, $M_\mathrm{tot}=2.3\times10^{15}\Msun$, this is still equal to around $2.17\%$. Using the method from Sec.~\ref{sec:method}, we find a total of 58 projections of galaxy clusters from the ``giants'' mass bin which manage to reproduce eight extremely massive substructures with a mass fraction $M_\mathrm{cyl}/M_\mathrm{tot}>2.17\%$. Note that $M_\mathrm{tot}$ is defined here as the total projected mass within $1.3\,\mathrm{Mpc}$ to be comparable to the results by \citet{jauzac16}. 

The case which best reproduced the measurements for Abell~2744 is a projection of galaxy cluster number~20 from the ``giants''. 
The central $(2\,\mathrm{Mpc})^2$ region of the best projection is shown in Fig.~\ref{fig:Cl20_A}. As seen on the top left, the eight most massive substructures (black circles) all lie close to the center (black x) and follow a strongly elongated distribution. For comparison the stellar surface density (bottom left) as well as a mock image in the r-band (bottom right) are shown. The black numbers are sorted by the mass of the substructures. Their mass as a fraction of $M_\mathrm{tot}$ is shown on the top right, with the values for this projection denoted as red crosses while the eight substructures observed in Abell~2744 by \citet{jauzac16} are shown as cyan diamonds. We find an exceedingly similar flat distribution of substructure mass fractions aside from the most massive one, in excellent agreement with the observations. This agreement can also be seen from the mass fractions of the individual substructures as shown in Tab.~\ref{tab:frac_mass}.
\begin{table}[!b]
    \centering
    \caption{The eight substructure masses determined within $150\,\mathrm{kpc}$~apertures for a projection of galaxy cluster~20 of the ``giants'' as well as Abell~2744, with values for the latter as determined by \citet{jauzac16}.}
    \label{tab:frac_mass}
    \begin{tabular}{|r|cccccccc|}
    \hline
    \rule{0pt}{0.9\normalbaselineskip} & {1} & {2} & {3} & {4} & {5} & {6} & {7} & {8} \\
    {ID} & {Core} & {NW} & {S3} & {N} & {S4} & {S2} & {W$_\mathrm{bis}$} & {S1} \\
    \hline
     {} & \multicolumn{8}{c|}{Mass fraction $[\%]$}\\
    \hline
    A2744 & 5.9 & 3.4 & 2.8 & 2.7 & 2.4 & 2.3 & 2.3 & 2.2 \\
    Cl20 & 3.6 & 3.5 & 2.6 & 2.6 & 2.6 & 2.3 & 2.2 & 2.2\\
    \hline
     {} & \multicolumn{8}{c|}{Total masses of structures $[10^{13}\Msun]$}\\
    \hline
    A2744 & 13.55 & 7.9 & 6.5 & 6.1 & 5.5 & 5.4 & 5.2 & 5.0  \\
    Cl20 & 3.7 & 3.7 & 2.7 & 2.7 & 2.7 & 2.4 & 2.3 & 2.3 \\
    \hline
    \end{tabular}
\end{table}

While this is the best fitting projection, it is not the only one, as can be seen from the shaded areas for each substructure in the upper right panel of Fig.~\ref{fig:Cl20_A}:
Underlayed in blue is the distribution of substructure masses for all~200 orientations of galaxy cluster~20, with the size of the bulge corresponding to the relative frequency of each value. The overall spread between orientations (blue vertical lines) is quite sizable, and it can be concluded that while the extreme mass distribution within Abell~2744 is reproducible it represents an outlier, with very high substructure mass fractions for the fifth to eighth substructures.

\begin{figure*}
  \begin{center}
  \includegraphics[width=0.95\textwidth]{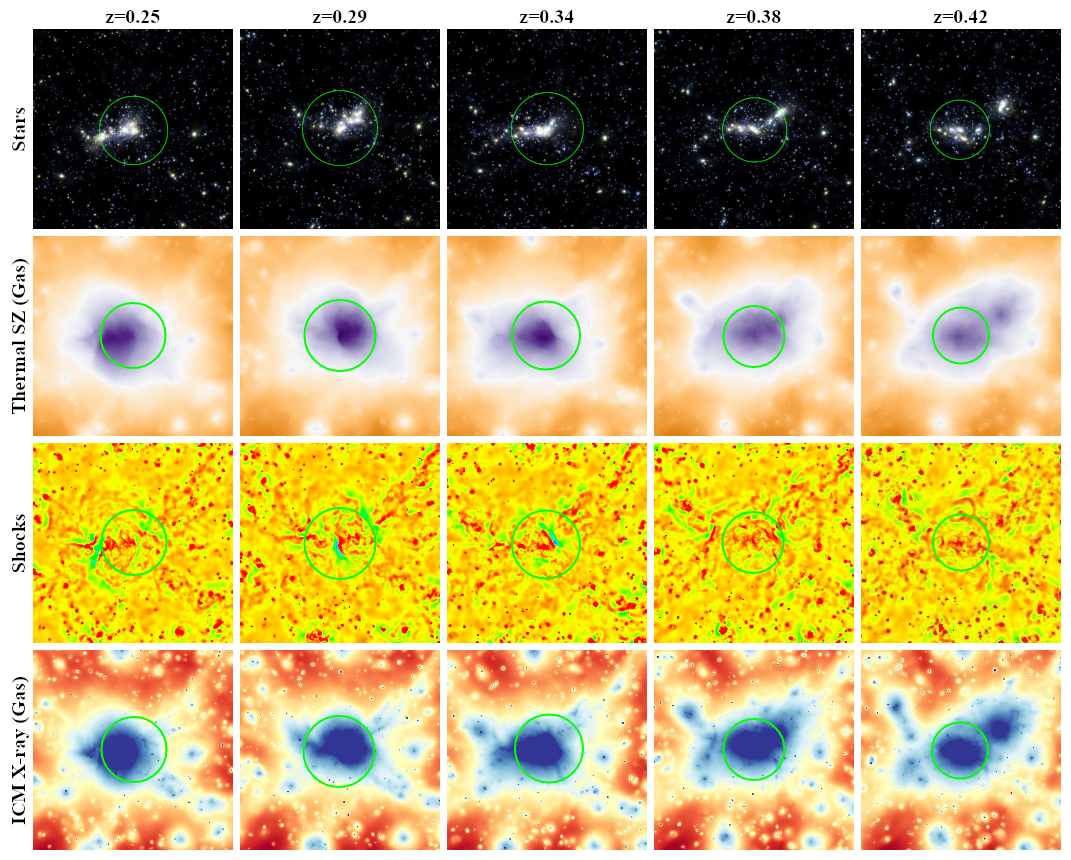}
  \caption{Redshift evolution of cluster~20 of the ``giants'', the Abell~2744 counterpart, with the different columns showing the different redshifts as indicated in the column titles. In all panels, the green circle marks the $R_\mathrm{500}$ radius of the cluster as calculated from the 3D mass distribution at each redshift. For scaling, at $z=0$ this corresponds to $R_\mathrm{500} = 1.94\mathrm{Mpc}$.
  \textit{Top row:} Stellar components.
  \textit{Second row:} Map of the thermal SZ-effect, shown as Compton-X maps, with color according to the thermal pressure from low (orange) to high (violet).
  \textit{Third row:} Shocks calculated from the Compton-X maps, with shocks in green or even blue if strongly compressed.
  \textit{Bottom row:} X-ray emission from the hot gas component, from low (red) to bright (dark blue).
  }
  {\label{fig:z_evo_20}}
  \end{center}
\end{figure*}

This means that the masses observed in Abell~2744 are reproducible within simulations also when including baryons, and arise from projection effects. The latter becomes especially clear when considering the orange points in the upper right panel of Fig.~\ref{fig:Cl20_A}, which mark the subhalo mass-fractions as determined by \textsc{SubFind}, i.e., when only counting the matter to the substructure that is really bound to it.
There only the second subhalo has a mass fraction comparable to the {\it eighth} substructure within Abell~2744, with all other subhalos lying far below. The mass fraction of the first subhalo lies much higher outside of the plotted range as it is the main halo and thus assigned all bound mass of the galaxy cluster not within other subhalos.
This clearly highlights the difficulty in comparing substructures obtained from simulation algorithms and observations, not just accounting for projection effects but also for the lack of possibility to separate bound from unbound structures in observations.

As outlined by \citet{jauzac16}, one of the eight measured substructures, W$_\mathrm{bis}$, is most likely a background source. When excluding the potential background source W$_\mathrm{bis}$, the number of projections of galaxy clusters from the ``giants'' mass bin with seven substructures of mass $M_\mathrm{cyl}/M_\mathrm{tot}>2.17\%$ rises to~108. However, the projections made here go out to a depth of $r_\mathrm{z}\approx 5\cdot r_\mathrm{vir}$. We can instead require all the substructures to not be background sources by projecting out only to a depth of $1\cdot r_\mathrm{vir}$. Therefore, we repeated the study for~200 orientations of the cluster which provided the best reproduction of Abell~2744, namely galaxy cluster~20. We find in this case~15 projections with seven substructures of mass $M_\mathrm{cyl}/M_\mathrm{tot}>2.17\%$ (and even~2 with eight), such that Abell~2744 is still well reproducible by galaxy cluster~20, even if only substructures inside the galaxy cluster virial radius are considered. A comparison between these different projection depths for ``giants'' clusters~5 and~20 is discussed in detail in Appendix~\ref{sec:depth}.

Note that the virial radius and mass of cluster~20 are $r_\mathrm{vir}=2.47\,\mathrm{Mpc}$ and $M_\mathrm{vir}=1.29\times10^{15}\Msun$, respectively, making it less massive overall than Abell~2744. For the best projection in particular, $M_\mathrm{tot}=1.03\times10^{15}\Msun$. Given the generally constant or increasing substructure mass fraction with increasing galaxy cluster mass from Fig.~\ref{fig:radial} and the self-similarity of the subhalo mass functions of all our galaxy cluster bins from Fig.~\ref{fig:subs}, it would stand to reason that a larger box would also be able to reproduce Abell~2744 with comparable absolute substructure masses.
This is supported also by the fact that galaxy clusters of a total mass comparable to Abell~2744 are found in the larger volume of the {\it Magneticum Pathfinder} simulation suite \citep[see Fig.~15 of][]{remus22_2}, however, the resolution of that simulation is to low for a substructure study as performed in this work.

The similarities in the substructure masses found for galaxy cluster~20 of the ``giants'' to Abell~2744 motivate a closer look at their origin. Fig.~\ref{fig:z_evo_20} shows the evolution of galaxy cluster~20 from $z=0.42$ (right column) to $z=0.25$ (left column), in maps depicting the stars, thermal SZ emission from the gas, the corresponding shocks, and finally the X-ray emission (top to bottom row). The images are centered on the deepest point of the potential at each time while the green circle denotes $r_\mathrm{500}$. 

\begin{figure}
  \begin{center}
  \includegraphics[width=0.975\columnwidth]{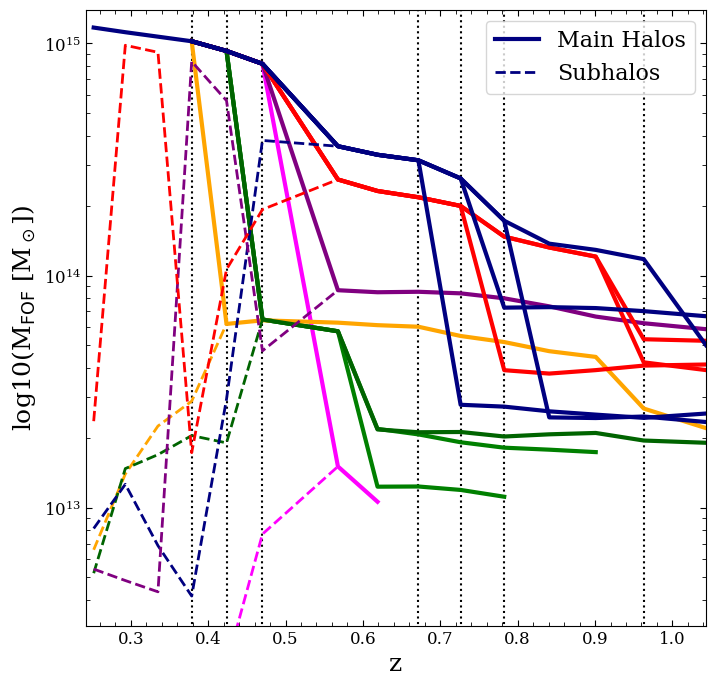}
  \caption{Merger history of Abell~2744 counterpart cluster~20 of the ``giants'', with main halos (solid) and their respective subhalos after infall (dashed lines). Subhalo masses are the summed mass of all bound particles. Only main halos with masses $M_\mathrm{FOF}\geq1\times10^{13}\Msun$ are shown. Branches belonging to the two components A and B of the major merger are in blue and red, respectively.
  }
  {\label{fig:merger_20}}
  \end{center}
\end{figure}

In the stellar maps we see that galaxy cluster~20 is the result of a recent major merger with a mass ratio of 1:1.4, with the initial two clusters visually distinct for $z=0.42$, where one is centrally located and the other to the upper right immediately outside $r_\mathrm{500}$ (henceforth cluster ``A'' and ``B''). B has been on an infalling trajectory since $z=0.67$ but has just reached a distance of $r_\mathrm{3d}=1.68\,\mathrm{Mpc}$ at $z=0.42$. Even though it is already within the $r_\mathrm{200}=2.15\,\mathrm{Mpc}$ of cluster A at this time \textsc{SubFind} still assigns it a mass of $M_\mathrm{sub,B}=1.1\times10^{14}\Msun$. Moving forward in time, cluster B falls within $r_\mathrm{500}$ of A at $z=0.38$, such that a significant portion of its mass is now assigned to cluster A, with $M_\mathrm{sub,B}$ dropping to $1.7\times10^{13}\Msun$. This coincides with a connection of regions of high X-ray emission from the hot gas halos as can be seen in the bottom row, going from two distinct peaks to one. 

Moving to $z=0.34$, the nearly radial infall of cluster B triggers a strongly peaked shock (see third row of Fig.~\ref{fig:z_evo_20}) to the immediate upper right of the cluster center of cluster A. As cluster B reemerges toward the bottom left at $z=0.29$ and then continues left until $z=0.25$, the shock front is pushed along. 
This is in excellent agreement with observations of shock fronts observed in Abell~2744 in X-ray with Chandra by \citet{owers11}, coinciding with radio relics \citep{eckert16,rajpurohit21}, that have been interpreted to result from a post core passage major merger with a resulting shock front toward the South-East \citep[see also][]{kempner04,boschin06}, exhibiting much the same characteristics found here for galaxy cluster~20.

Following the evolution of Cluster~20 further back in time as shown in Fig.~\ref{fig:merger_20}, we find that it actually only assembled to a mass above $M_\mathrm{FOF} \geq 1\times10^{14}\Msun$ at about $z\approx0.96$, with multiple group merger events happening simultaneously: between $z=0.62$ and $z=0.42$, four halos with masses above group mass (i.e., $M_\mathrm{FOF} \geq 1\times10^{13}\Msun$) are accreted onto the main halo A in addition to cluster B, with mass ratios relative to Cluster A at the time of merging of about 1:4, 1:6, 1:6, and 1:24. Note here that, different than for galaxies, a 1:24 merger is still a massive merger event in case of galaxy clusters, as this is a group being accreted onto the cluster which, by itself, already harbors several galaxies and a hot gaseous halo. This supports claims from observations that Abell~2744 actually not only originated from a single major merger but rather several ongoing multiple merger events \citep[e.g.,][]{merten11,rajpurohit21}. Indeed, we find that of the total growth in mass from $z=0.67$ to $z=0.25$ $58\%$ comes from mergers with halos of mass larger than $1\times10^{13}\Msun$. This means that for the case of Cluster~20 a significant amount of the mass of the cluster is pre-processed in groups. 

Additionally, the lower row of Fig.~\ref{fig:z_evo_20} clearly shows that cluster~20 is fed through four filaments, especially well visible at $z=0.29$ and $z=0.34$. This supports the idea that the multiple filaments observed by \citet{eckert15} in the hot gas map constructed for Abell~2744 using XMM-Newton data could be depicting four feeding filaments surrounding the cluster.
All this gives further evidence that Cluster~20 from the Magneticum simulation indeed resembles Abell~2744. We conclude that Abell~2744 is indeed a rare case of a violently assembling galaxy cluster, sitting at a node in the cosmic web that only recently started to collapse from multiple different directions.

\subsection{Accretion and Dynamical State}\label{subsec:accdyn}
\begin{figure*}
  \begin{center}
  \includegraphics[width=0.95\textwidth]{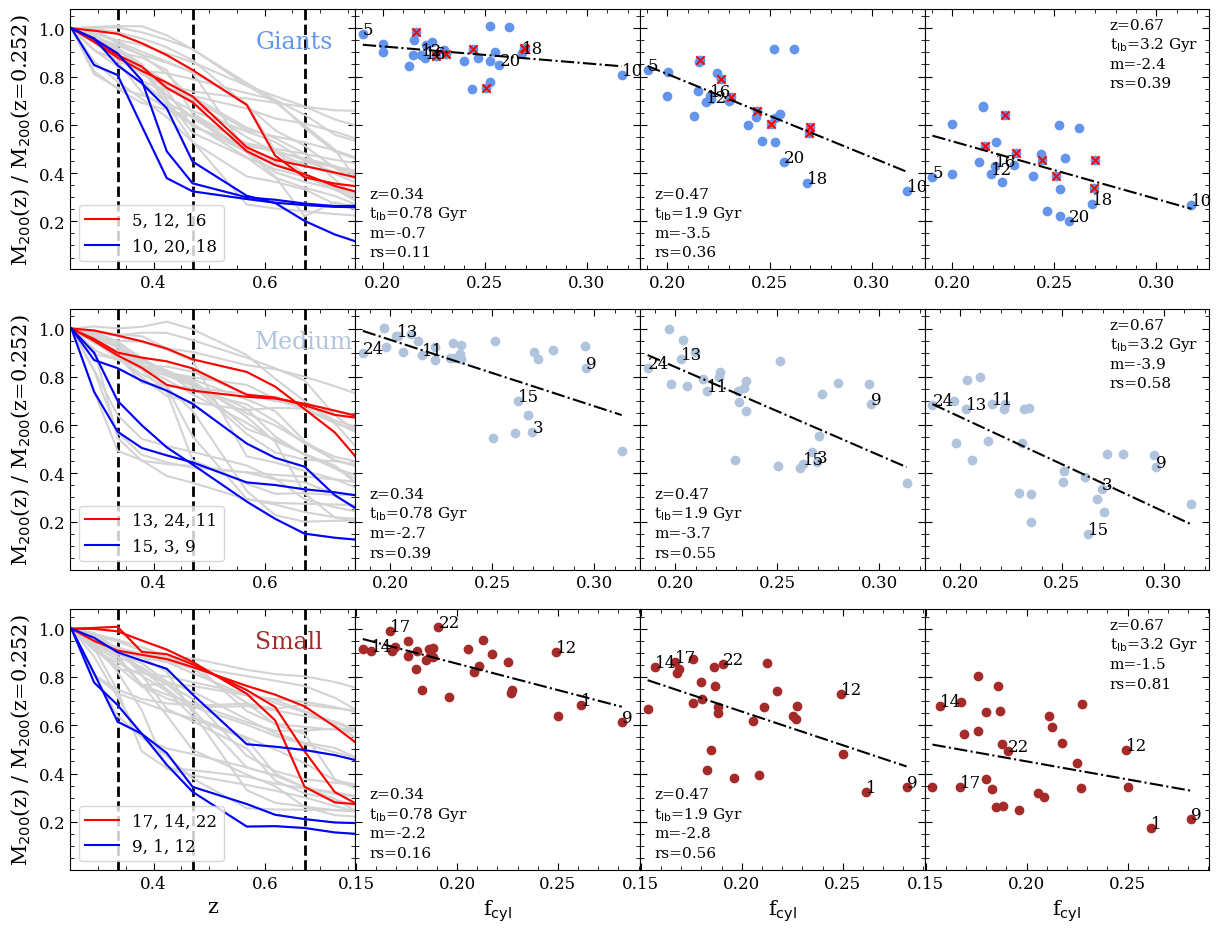}
  \caption{The mass history of the galaxy clusters of the ``giants'' to the ``small'' mass bins (top to bottom row) as a function of redshift ({\it first column}). The three galaxy clusters with the highest (lowest) $f_\mathrm{8}$ are highlighted in blue (red), with their numbers given in the legend, and the three vertical black lines indicate three redshifts ($z=0.34,0.47,0.67$). For each the mass fraction of the final mass of the clusters is depicted as a function of $f_\mathrm{cyl}$ ({\it second to fourth column}), and the black dash-dotted line is the best fit to the points. The slope $m$ and residual $rs$ of the fit are given in the legend. $t_\mathrm{lb}$ is the lookback time from the final redshift $z=0.252$. Numbers indicate the six highlighted clusters, while the red crosses for the ``giants'' indicate galaxy clusters which were classified as protoclusters at $z=4$ by \citet{remus22_2}.}
  {\label{fig:acc}}
  \end{center}
\end{figure*}

Given the flat distribution of the mass fractions of the substructures for both Abell~2744 and our Cluster~20 (see upper right panel of Fig.~\ref{fig:Cl20_A}, the mass fraction of the eighth substructure is still rather large. As we have seen, our best matching cluster to Abell~2744, cluster~20, is highly dynamical active and dominated by recent multiple accretion events. Thus, the question arises how well the mass fraction of the eighth substructure would be in tracing the recent assembly history of a cluster, independent of the galaxy cluster mass.

To this end, we calculate for each cluster in our four cluster mass bins the median mass fraction of the eighth substructure from all projections, $f_\mathrm{8}$.
The first column of Fig.~\ref{fig:acc} depicts the mass accretion histories of the galaxy clusters of the ``giants'', ``medium'', and ``small'' cluster mass bins, repsectively. More explicit, the mass fraction of the final cluster mass (at $z=0.252$) is shown as a function of redshift. 
All individual clusters are shown as gray lines, with the three clusters with the highest $f_\mathrm{8}$ for each cluster mass bin shown in blue, and those three with the lowest $f_\mathrm{8}$ marked in red.

As can be seen immediately, for all cluster mass bins the clusters with the lowest $f_\mathrm{8}$ show relatively flat recent accretion histories, with only small amounts of mass accreted in at least the last $2\,\mathrm{Gyr}$, while all clusters with the highest $f_\mathrm{8}$ have accreted at least half of their mass in the last 2~to~$3\,\mathrm{Gyr}$. More specific, for the ``giants'' we find that the three galaxy clusters with the highest $f_\mathrm{8}$ have gained around 55~to~$68\%$ of their final mass starting from $z\approx0.47$ (equaling around $2\,\mathrm{Gyr}$), compared to 17~to~$31\%$ for those with the lowest $f_\mathrm{8}$.
This clearly shows that the relative mass of the eighth substructure is a good indicator for the amount of accretion in the last $2\,\mathrm{Gyr}$, as large values in $f_\mathrm{8}$ can only be reached through recent large mass infall with a high likelyhood for a major merger event.

Given that the mass fraction of the eighth substructure is in first order an arbitrary choice, we also tested the correlation between the recent mass accretion history with the mass fractions of the fourth, fifth, and sixteenth substructure mass fractions $f_\mathrm{4}$, $f_\mathrm{5}$, and $f_\mathrm{16}$. While the split is surprisingly the clearest for the eighth substructure fraction $f_\mathrm{8}$, both $f_\mathrm{4}$ and $f_\mathrm{5}$ give similarly good results, while the signal vanishes when using the sixteenth substructure mass fraction. We also inspected the properties found for the second most massive substructure, however, the connection to the mass history was less clear, indicating that a single merger in the recent accretion history of a cluster leading to a single massive remaining substructure does not necessarily trace a violent multiple merger dominated mass accretion history but is simply an occurring event for any cluster \citep[see][for more details on the mass ratio of the second most massive substructure and the so-called fossilness parameter]{ragagnin19}. 

To avoid picking a single substructure to trace the mass accretion history, one can also use the total mass fraction contained within substructures $f_\mathrm{cyl}$ as a tracer for the dynamical state of a galaxy cluster. Previous studies found that higher $f_\mathrm{cyl}$ indicate a dynamically active galaxy cluster \citep{delucia04,neto07,biffi16}. This is because once a subhalo falls into a larger halos influence, it begins being disrupted by processes such as ram-pressure and tidal stripping. As the accretion of new subhalos is slower than the disruption at lower redshifts \citep{jiang16}, this leads to a net reduction of $f_\mathrm{cyl}$ with time and consequently a higher $f_\mathrm{cyl}$ correlates with a more recent formation time as described by \citet{jiang17}.

While using $f_\mathrm{cyl}$ is not as efficient as $f_\mathrm{8}$ in identifying the extremes of the dynamically active or passive clusters, we find it to be an overall excellent tracer for the recent mass accretion history.
To more closely quantify the relation between final substructure mass and the mass accretion history, we select three redshifts as indicated by the vertical black lines in the left panels of Fig.~\ref{fig:acc} and plot the mass fractions of the final mass of the clusters at these redshifts as a function of the final substructure mass fraction $f_\mathrm{cyl}$ in the three right columns of Fig.~\ref{fig:acc}. 

For the first at $z=0.34$ all clusters have only accreted little mass, with a negligible dependence on $f_\mathrm{cyl}$. This can also be seen from the shallow slope $m$ of the best-fit line (black dash-dotted) given in the legend. Going further back to $z=0.47$, however, shows a steep trend toward decreasing mass fraction with increasing $f_\mathrm{cyl}$, which means that galaxy clusters exhibiting high substructure masses are indicative of a large amount of recent (within $2\,\mathrm{Gyr}$) mass accretion. This trend persists (although with a more shallow slope) out to $z=0.67$, with a comparable scatter as indicated by the residual of the fit (given as $rs$). 

The same correlation between the total mass fraction in substructures $f_\mathrm{cyl}$ and the cluster mass accretion history can be found for all galaxy cluster mass bins as shown in Fig.~\ref{fig:acc} (``giants'': upper row, ``medium'': middle row, and ``small'': bottom row), with two notable exceptions. First, the less massive galaxy clusters already show a noticeable slope at $z=0.34$, though this is likely due to none of the clusters of the ``giants'' having accreted as high a mass fraction within the same period. Secondly, the scatter in the relation is higher at $z=0.67$ for the ``medium'' and ``small'' galaxy clusters compared to the ``giants'', as can be seen from the fit residual. Nonetheless, we still find a downwards slope such that the final substructure mass fraction measured in projection $f_\mathrm{cyl}$ is still a tracer for the mass accretion history, even out to a lookback time of $t_\mathrm{lb}=3.2\,\mathrm{Gyr}$. This time period is larger than the mass-loss timescale of subhalos as given by \citet{jiang17} for dark matter only halos, which may indicate that the presence of baryons allows the signal to persist for longer.
We conclude that a large substructure mass fraction $f_\mathrm{cyl}$ is an excellent tracer for a large amount of accretion within the last $2\,\mathrm{Gyr}$ and can be indicative of recent major merger events, with really large values of $f_\mathrm{cyl}$ even hinting at more merger events happening at even earlier times, without the corresponding stellar cores of these structures being disrupted.
\begin{figure*}
  \begin{center}
  \includegraphics[width=0.98\textwidth]{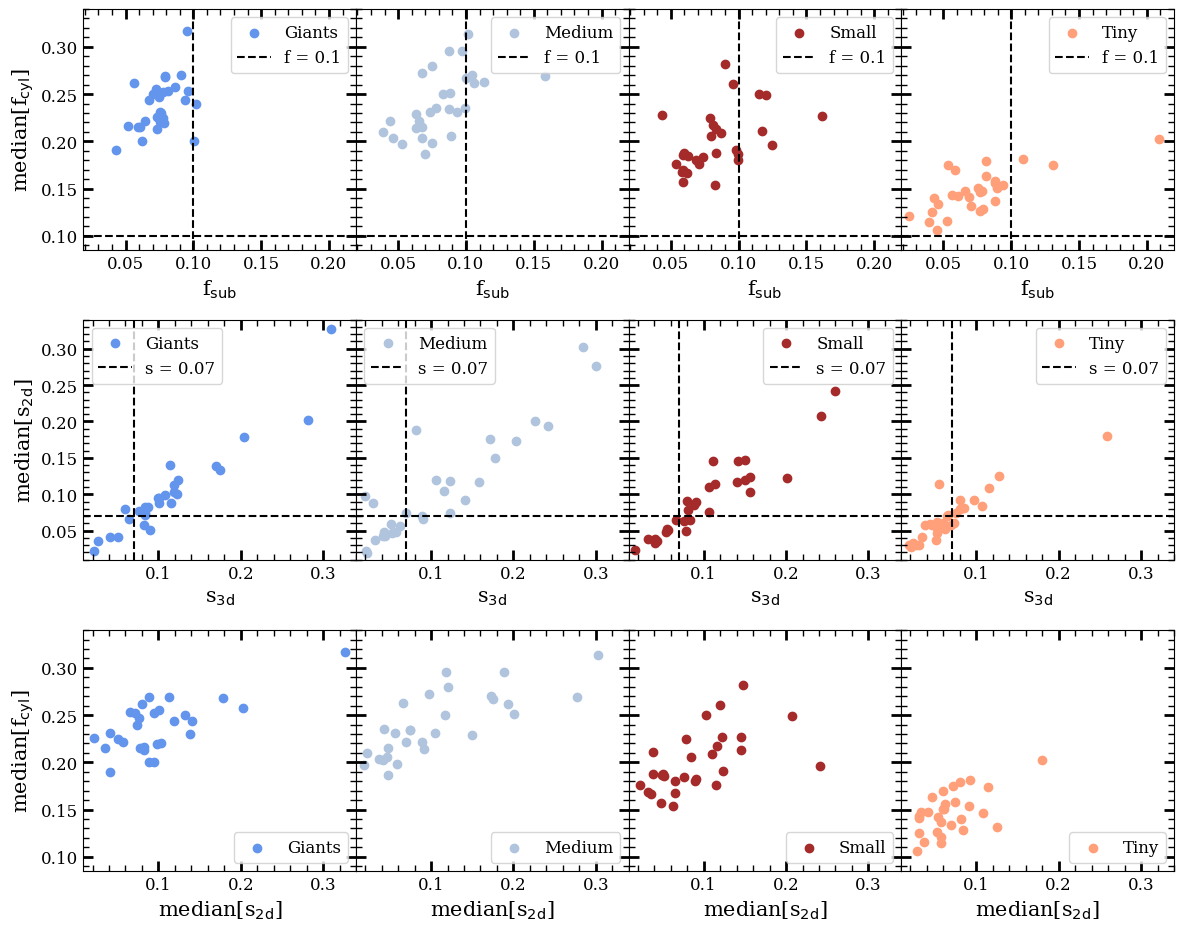}
  \caption{Median projected values of $f$ ({\it top row}) and $s$ ({\it middle row}) over the~200 random projections for each galaxy cluster as a function of their three-dimensional values, as well as the relation between projected $f$ and $s$ ({\it bottom row}). The dashed lines denote typical thresholds, with higher values indicative of a dynamically active galaxy cluster.}
  {\label{fig:dyn}}
  \end{center}
\end{figure*}

Additionally, the red crosses in Fig.~\ref{fig:acc} at all three redshifts of the ``giants'' indicate galaxy clusters which were classified as protoclusters at $z=4.2$ by \citet{remus22_2}, that is clusters for which their progenitor had assembled at least a mass of $M_\mathrm{vir} \ge 1\times10^{13}M_\odot$ already at $z=4.2$, i.e., they are nodes that started collapsing early-on. Interestingly, fulfilling certain conditions to be a protocluster at high redshifts (for example having a high star-formation rate, high number of substructures or being very massive) is found here not to correlate either with a particularly high or low resulting substructure mass fraction $f_\mathrm{cyl}$, nor with noticeable deviations from the best-fit lines. Instead, those clusters that started collapsing early behave here comparably to those which did not yet start to collapse and build up into a protocluster at $z\approx4$.
This clearly indicates that neither the mass nor the activity of assembly (i.e., the substructure mass fraction $f_\mathrm{cyl}$) observed at low redshifts is an indicator for the cluster having assembled a significant amount of its mass already at high redshifts. In fact, this is in agreement with our results for Cluster~20, the Abell~2744 analogue, that we found to have assembled only recently below $z=1$, while it at the same time is one of the most massive galaxy clusters in our simulation volume, as well with the fact that only about $25\%$ of our most massive galaxy clusters are identified as protoclusters at redshifts of $z\approx4$.
Nevertheless, we do find one trend for the seven protoclusters present in out sample: the four protoclusters selected by their high number of substructures at $z\approx4$ are also those which result in a higher $f_\mathrm{cyl}$ compared to the other three selected by mass or star-formation rate.
This could indicate that these particular clusters sit at big nodes of the cosmic web, building up into future super clusters, however, we note that the sample here is of low significance due to the small number of clusters.

\subsection{Substructures and Centershift}\label{subsec:cshft}
\begin{figure*}
  \begin{center}
  \includegraphics[width=0.98\textwidth]{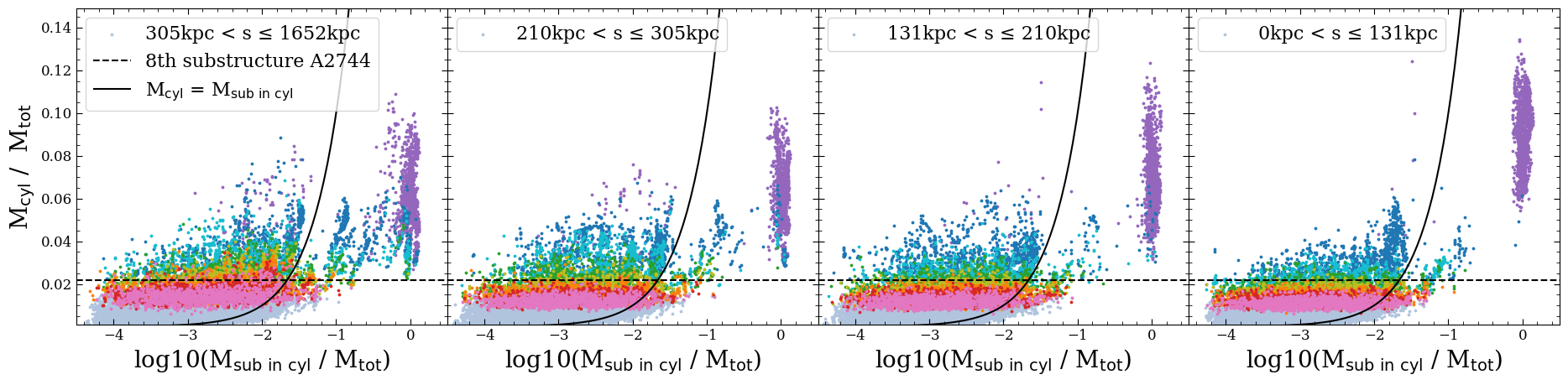}
  \caption{The substructure mass divided by the total mass within $1.3\,\mathrm{Mpc}$ in projected as a function of the most massive bound subhalo mass $M_\mathrm{s}$ within their aperture. The projections are split into quartiles by their centershift (each containing~1450), going from \emph{left} to \emph{right} as highest to lowest with borders as given in the legends. Within each projection, the first to eighth most massive substructures are colored as purple, blue, cyan, green, olive, orange, red and pink, with all others colored gray. The horizontal black line denotes the mass fraction of the eighth substructure of Abell~2744, such that all pink points which lie above it belong to a projection with a comparable substructure mass distribution. This is the case for $(57,1,0,0)$ projections when going from left to right. The black dashed curve denotes equality between projected and bound mass.}
  {\label{fig:centershift}}
  \end{center}
\end{figure*}
Another parameter which according to the literature correlates to the dynamical state of a galaxy cluster is the centershift $s$, that is the distance between the point of lowest potential (highest density) to the center-of-mass \citep{biffi16}. As with $f_\mathrm{cyl}$, a higher centershift $s$ indicates a dynamically active galaxy cluster, and the absolute value is typically divided by the virial radius to allow for comparisons between galaxy clusters of varying masses and sizes. Generally, thresholds of $f>0.1$ and $s>0.07$ are defined to be indicative of dynamical activity \citep{neto07}. 
However, as we have seen the large differences between the total substructure mass fraction in projection $f_\mathrm{cyl}$ and the total subhalo mass fraction in three-dimensions $f_\mathrm{sub}$ found in Fig.~\ref{fig:mcylmvir} cause problems when consistently classifying the dynamical state in observations and simulations, and thus for the substructure mass fractions projection effects and the neglect of a binding criterion need to be considered.
Thus, in the following we will investigate the importance of such projection effects and the neglect of a binding criterion as used in three-dimensional substructure identifications for using $f$ and $s$ as tracers of dynamical activity.

Fig.~\ref{fig:dyn} shows in the first two rows the median projected values of $f$ and $s$ as a function of their three-dimensional values, with the thresholds for dynamical activity from \citet{neto07} given as dashed lines. Although we find a trend to increasing $f_\mathrm{cyl}$ with increasing $f_\mathrm{sub}$ on the top row, the projected $f_\mathrm{cyl}$ all lie above the threshold even when their $f_\mathrm{sub}$ are lower than the threshold. This would necessitate either an increased threshold in projection or a decreased threshold in three dimensions to somewhat agree when predicting dynamical activity.  

Conversely, the centershift is found to agree well between projection and three dimensions. Overall, the number of clusters whose dynamical state would be classified the same between two to three dimensions is found here to be~18 of~116 when using $f$ as a tracer, compared to~103 of~116 when using $s$ (or around $15.6\%$ versus $88.8\%$).

Given that the centershift is a consistent determinant for the dynamical state independent of whether it is measured from projections in observations or intrinsically from simulations, the question is how well it correlates with $f_\mathrm{cyl}$. We indeed find a good agreement in that high centershift corresponds to high substructure masses, as shown in the bottom row of Fig.~\ref{fig:dyn}, though there exists a noticeable scatter.

Nonetheless, the found correlation prompts the question of the impact of a high centershift on {\it individual} substructure masses for a given projection. To this end, the projections of the galaxy clusters within the giants mass bin are split into four groups based on their centershift. Then, the resulting fractional projected substructure masses $f_\mathrm{x}=M_\mathrm{cyl}/M_\mathrm{tot}$ are depicted in Fig.~\ref{fig:centershift} as a function of the most massive bound subhalo mass fraction $M_\mathrm{sub}/M_\mathrm{tot}$ within their aperture. The colors denote the different ranks of the substructures, from $f_\mathrm{1}$ the first substructure shown in purple to the eights most massive substructure $f_\mathrm{8}$ shown in pink.

There arises a trend, with the mass fractions $f_\mathrm{1}$ of the first substructures generally increasing with decreasing centershift while the mass fractions of the second to eighth substructure generally decrease. For high centershifts $s$ there are multiple cases where the most massive substructure (purple) lies at a low value of $M_\mathrm{s}$ (on the order of $1\%\cdot M_\mathrm{tot}$), while this is exceedingly rare for low centershifts $s$, where instead the most massive substructure also typically lies at the deepest point of the potential (thus containing the \textsc{SubFind} center of the cluster). 

The horizontal black lines in Fig.~\ref{fig:centershift} denotes the mass fraction $f_\mathrm{8}$ of the eighth substructure of Abell~2744. For the lowest centershift bin there are only~23 of~1450 projections with a fifth substructure (olive) where the substructure mass fraction $f_\mathrm{5}$ lies above this black line, and not even a sixth substructure reaches a fraction comparable to the eigths substructure mass fraction of Abell~2744. For the highest centershift bin, this occurs more frequently even for the eighth substructure (pink), with~57 of~1450 projections having an eighth substructure with a mass fraction $f_\mathrm{8}$ larger or comparable to Abell~2744. There is only one projection with an eighth substructure mass fraction $f_\mathrm{8}$ above the line which does not lie in the bin of highest centershift, and it instead lies in the bin with second highest. We conclude that multiple massive substructures can only be found in non-relaxed clusters with a large centershift, even if in return a large centershift does not necessarily indicate a large number of massive substructures. This is because a single major merger which is not yet relaxed can also lead to a large centershift, but may result in only two massive substructures.

We thus find that a high centershift enhances the likelihood to find a larger number of massive substructures. This can be understood when considering the relation of the line-of-sight to the feeding filaments or to the axis of a recent major merger. For the galaxy clusters where there is a dominant such axis, if one were to look straight along it the substructures would appear clumped together and thus result in a lower centershift. This would also reduce the number of very massive substructures as multiple of them may be projected into single apertures. Conversely, orientations looking at the axis perpendicularly will see a string of individual substructures sitting within a strongly elongated halo, thus allowing a higher number of apertures with extreme masses, while the center of these mass clumps is somewhere in between, shifted strongly compared to the potential minimum which will be inside the most massive of these substructures. Indeed, \citet{eckert15} find indications that the merger axis for Abell~2744 is perpendicular to the line-of-sight, comparable to what we find here for galaxy cluster~20 of the ``giants'' as discussed in Sec.~\ref{subsec:abell}.

\section{Summary and Conclusion}\label{sec:summary}
In this study we aimed at comparing substructure mass distributions of simulated galaxy clusters to those observed with gravitational lensing, as some of the observed galaxy clusters, especially Abell~2744, exhibit such massive substructures that these were discussed to possibly be in tension with results from cosmological $\Lambda$CDM simulations. So far, previous studies have tried to solve the tension using dark matter only simulations. Here, for the first time, we used a fully hydrodynamical cosmological simulation to search for Abell~2744 counterparts and study the substructure mass functions of galaxy clusters in projection.

As we aim at comparing to some of the most massive structures in the known Universe, a large volume simulation with proper resolution is required. To this end, we use the {\it Box2b/hr} of the hydrodynamical cosmological simulation suite {\it Magneticum Pathfinder}, one of the largest baryonic simulations currently available covering a volume of (909 cMpc)$^3$. In this simulation volume, at $z=0.252$ there are 29 galaxy clusters with masses above $M_\mathrm{FOF} \geq 10^{15}M_\odot$, and when requiring halos to contain at least~100 stellar particles galaxies are resolved down to stellar masses $M_* \geq 5\times10^{9}M_\odot$. While having more than 500 of such resolved member galaxies in the most massive cluster, this large dynamical range allows us to study galaxy cluster substructure properties down to $M_\mathrm{vir} \geq 10^{14}M_\odot$ with large significance.

We developed a procedure for determining substructure masses within galaxy clusters in projection, but after subtracting a spherical model for the cluster, which more closely follows the procedure typically applied to observations. The properties of the substructures found through this procedure were then compared to those of the subhalos, directly determined from the three-dimensional particle distribution by the structure finder \textsc{SubFind} commonly used to find subhalos in simulations, as well as to the observations of especially the extreme case of galaxy cluster Abell~2744.

Using this method, we find that:
\begin{itemize}
    \item the total projected substructure mass generally is a factor of two to three larger than the total bound intrinsic subhalo mass. For some galaxy clusters, the spread in projected substructure mass fractions, $f_\mathrm{cyl}$, varies by a factor of four depending on the projection.
    \item the contributions to the substructure masses from the residuals of the main halo even after subtracting a spherical model are more significant than those from additional subhalos projected in a line. We quantified this into a two-parameter model which for a given subhalo mass and projected distance for a given host galaxy cluster mass can produce an expected range of projected substructure masses within a factor of~2 for around $80\%$ of the cases.
    \item we can successfully reproduce the substructure mass fractions observed within Abell~2744 by \citet{jauzac16}. Nonetheless, it constitutes a rare projection. Of the~5800 total projections of galaxy clusters with mass above $1\times10^{15}\Msun$, only~58 (108) have eight (seven) substructures with mass fractions comparable to the least massive substructure found in Abell~2744.
    \item our best reproduction of the substructure mass fractions of Abell~2744 occurs for a galaxy cluster that just recently underwent a massive major merger event with a merger ratio of 1:1.4, in addition to several minor merger events with mass ratios of 1:4, 1:6, and 1:24. We find that the appearance of multiple large substructures is a direct consequence of such recent multiple merger accretion events.
    \item furthermore, our simulated counterpart of Abell~2744 exhibits strong noticeable shock fronts resulting from the post core passage major merger, in line with observations of strong shock fronts detected in Abell~2744 \citep[e.g.,][]{owers11,rajpurohit21}. In addition, we find that the simulated counterpart of Abell~2744 is fed through four main filaments, again in agreement with observations \citep{eckert15}.
    \item in general, a large total substructure mass fraction correlates to a larger amount of recently accreted mass and the dynamical state of the cluster, in agreement with previous works by \citet{neto07,jiang17}. For the galaxy clusters with $M_\mathrm{vir}\geq2\times10^{14}\Msun$ this correlation persists even out to $3.2\,\mathrm{Gyr}$, though it becomes quite scattered for the lower mass clusters. This clearly indicates that the appearance of multiple massive substructures, independent of the host cluster mass, is a good tracer for (multiple) massive merger events occurring within the last $3.2\,\mathrm{Gyr}$, not yet long enough ago to completely disrupt the substructures due to stripping and dynamical friction processes in the cluster environment. This timescale is longer than what was previously found from dark matter only simulations, showing that the deeper potential wells generated due to the presence of the baryons fosters the longer survival times of the satellite galaxies \citep[see also][for satellite galaxy survival timescales]{bahe19}. Ultimately, this increases the time window of the appearance of large substructures and therefore allows for more extreme configurations of total substructure mass fractions in hydrodynamical simulations compared to dark matter only ones.
    \item galaxy clusters whose eighth substructures have the highest (lowest) mass fractions out of our sample have gained a large (low) percentage of their mass within the last $2\,\mathrm{Gyr}$, allowing for insight into the dynamical state and recent accretion history of the galaxy cluster without the requirement of having measured all substructure masses.
    \item the mass of the observed galaxy cluster is not necessarily a guarantee that a significant fraction of the current cluster has been assembling already at high redshift, in agreement with results from protocluster evolution pathways by \citet{remus22_2}. Instead, we find some of the most massive galaxy clusters to have reached masses above $1\times10^{14}M_\odot$ only recently, below $z=1$, clearly demonstrating that some of the most massive nodes in the cosmic web have been assembled in very complex ways, being the result of larger numbers of smaller systems merging together on comparatively shorter timescales. Multiple massive substructures might be a good indicator for such nodes, however, the present sample of massive galaxy clusters is too small to provide statistically representative conclusions on that matter.
    \item the necessary condition for finding a large number of massive substructure masses is a high centershift. This means that the galaxy cluster must be dynamically active {\it as well as} observed from a projection angle where the substructures are distributed largely perpendicular to the line-of-sight. Such in-plane geometry then also fosters the detectable appearance of other indications of dynamically strongly disturbed systems like the presence of radio relics.
\end{itemize}
We conclude that, given the importance of galaxy cluster substructures for quantifying the clusters dynamical state as well as for constraining models of dark matter \citep{bhatta21}, it is of great importance to consider in detail the impact of projection effects when comparing measurements to cosmological simulations. The findings here further demonstrate that the total substructure mass fraction $f_\mathrm{cyl}$, even when measured in projection,
indicates dynamically active galaxy clusters and still correlates well with the mass accretion history of the cluster.

Finally, we have shown that we can successfully reproduce not just the substructure mass fractions but also other properties of the extreme galaxy cluster Abell~2744 in the hydrodynamical cosmological {\it Magneticum Pathfinder} simulation within the $\Lambda$CDM framework, demonstrating that there is no tension between $\Lambda$CDM and the existence of such massive substructures. Instead, we showed that such special galaxy clusters are rather interesting in terms of their accretion histories, and that they could be tracing special nodes of particularly late assembly in the cosmic web which, however, remains to be further analyzed in future studies.

\begin{acknowledgements}
LCK and KD acknowledge support by the COMPLEX project from the European Research Council (ERC) under the European Union’s Horizon 2020 research and innovation program grant agreement ERC-2019-AdG 882679.
The {\it Magneticum} simulations were performed at the Leibniz-Rechenzentrum with CPU time assigned to the Project {\it pr83li}. This work was supported by the Deutsche Forschungsgemeinschaft (DFG, German Research Foundation) under Germany's Excellence Strategy - EXC-2094 - 390783311. We are especially grateful for the support by M. Petkova through the Computational Center for Particle and Astrophysics (C2PAP). Plotting is done via matplotlib by \citet{hunter07}, and Gaussian fitting performed using the JULIA ``Distributions.jl'' package by \citet{besancon21}.

\end{acknowledgements}

\appendix

\section{The Gaussian Distribution of Substructure Masses}\label{sec:gauss}

\begin{figure*}
  \begin{center}
  \includegraphics[width=0.95\textwidth]{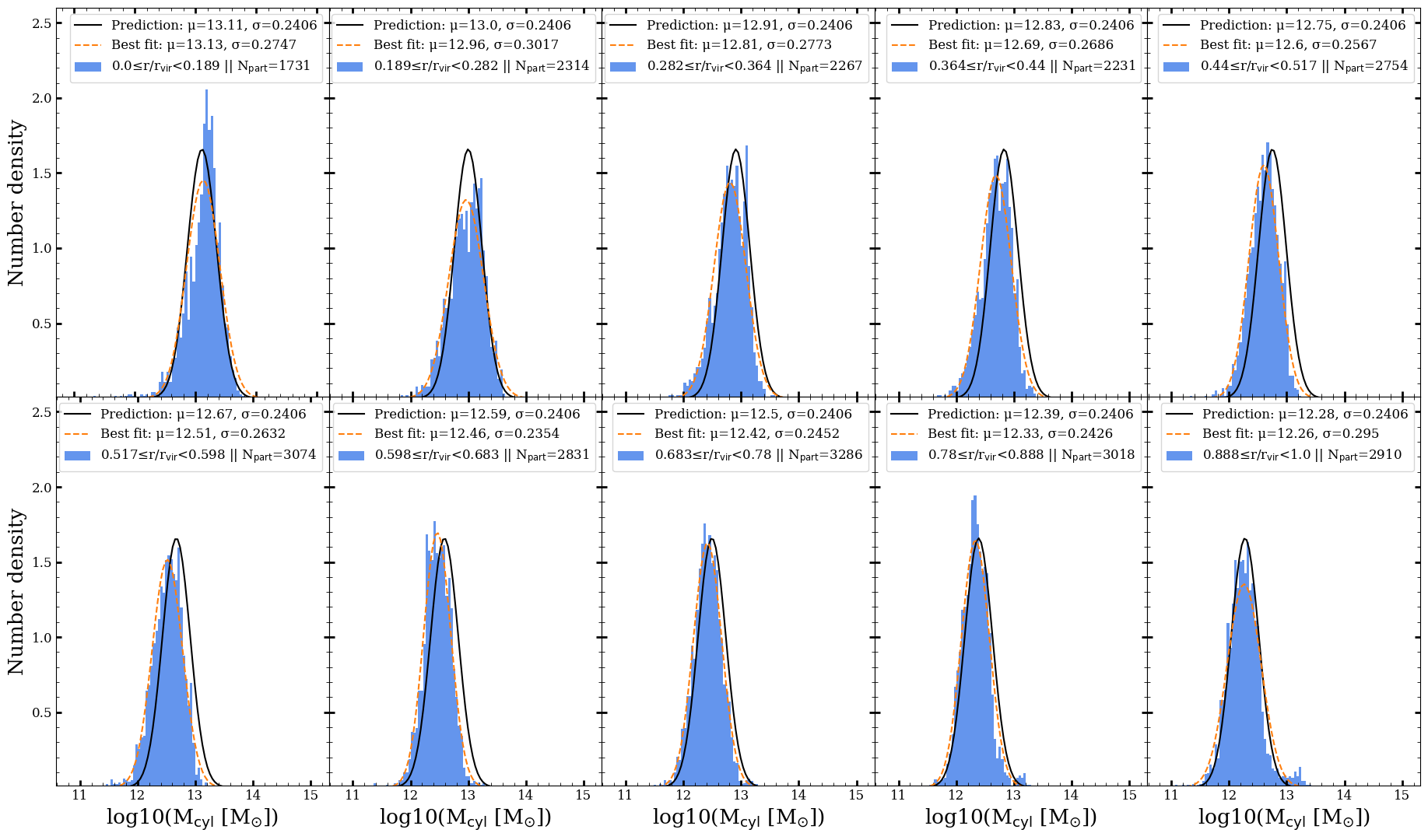}
  \caption{The distribution of substructure masses $M_\mathrm{cyl}$ for an example bin in subhalo mass, $6\times10^{11}\Msun\leq M_\mathrm{s}<8\times10^{11}\Msun$. Plotted are different ranges of $r_\mathrm{2d}$ as written in the legend going from lowest to highest projected distance as {\it left} to {\it right}, then {\it top} to {\it bottom}. The blue histograms depict the distribution of determined $M_\mathrm{cyl}$. Black solid lines are the predicted distributions of projected mass from Tab.~\ref{tab:para} via Eq.~\ref{eq:fit} while the orange dashed lines are the best-fit single Gaussian models to the $M_\mathrm{cyl}$, with mean and variance as given in the legend for both.}
  {\label{fig:mixture}}
  \end{center}
\end{figure*}

To see the distribution of substructure masses within a given subhalo mass bin $M_\mathrm{s}$, Fig.~\ref{fig:mixture} considers all substructures whose most massive bound subhalo has a mass of $6\times10^{11}\Msun\leq M_\mathrm{s}<8\times10^{11}\Msun$. This mass bin is split then further into ten radial bins by their projected distance. The resulting distribution of $M_\mathrm{cyl}$ (blue histogram) in each bin is largely Gaussian in nature, which can be seen from the direct fits (orange dashed curves). Using the mean mass $\bar{M}_\mathrm{s}=7.1\times10^{11}\Msun$ and the mean radius $\bar{r}_\mathrm{2d}$ of each radial bin allows a predicted distribution of $M_\mathrm{cyl}$ to be made via Eq.~\ref{eq:sigma} and Eq.~\ref{eq:fit}, which is shown as black curves. As can be seen, the prediction well captures the real distribution of substructure masses.

\section{The Impact of Varying Projection Depth}\label{sec:depth}

\begin{figure}
  \begin{center}
  \includegraphics[width=0.85\textwidth]{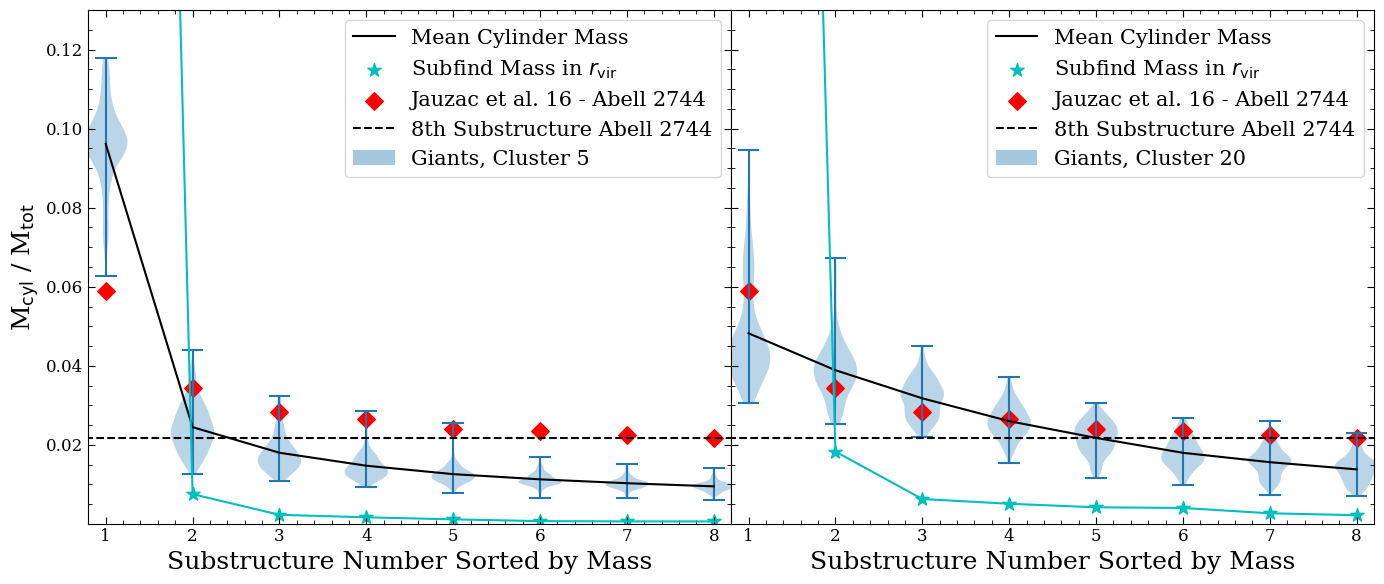}
  \includegraphics[width=0.85\textwidth]{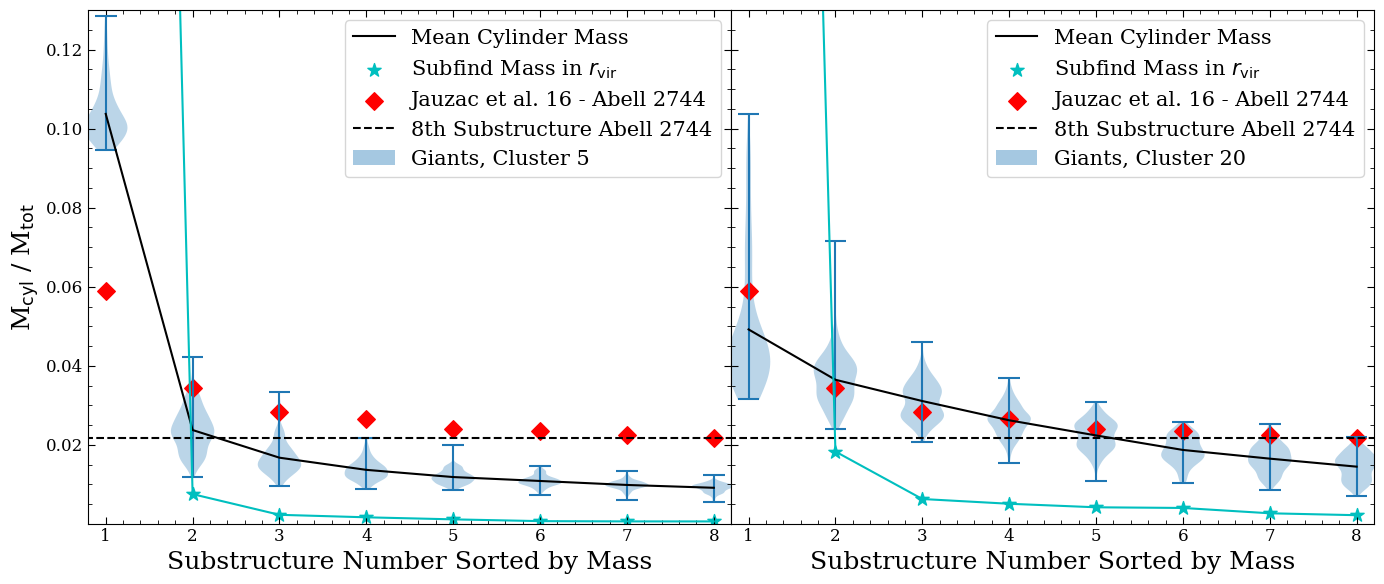}
  \caption{The distribution of the mass fractions for the eight most massive substructures within $1.3\,\mathrm{Mpc}$ for galaxy clusters~5 ({\it left}) and~20 ({\it right}) of the ``giants'' mass bin for all~200 orientations out to the typical projection depth $r_\mathrm{z}$ ({\it top row}) and out to $r_\mathrm{vir}$ ({\it bottom row}). Blue error bars denote the overall spread while the blue shaded area represents the percent number of orientations at each mass fraction, with a larger bulge meaning a larger percentage. Dashed black line equals a mass fraction of around $2.2\%$.}
  {\label{fig:r_5-20}}
  \end{center}
\end{figure}

To determine the impact of varying projection depths on the results presented, Fig.~\ref{fig:r_5-20} considers for two different projection depths two galaxy clusters, number~5 (left) and~20 (right) of the ''giants" mass bin. They are chosen as they represent a highly relaxed and disturbed cluster, respectively (see for example Fig.~\ref{fig:method}). Their fractional substructure mass for their eight most massive ones is depicted through a violin plot, with the blue error bars denoting the overall spread in the~200 projections while the frequency of mass fractions for each substructure number is represented through the blue shaded area. The black line denotes the mean mass over all orientations. First the top row with the typical projection depth $r_\mathrm{z}\approx 5\cdot r_\mathrm{vir}$ is considered.

Galaxy cluster~5 exhibits a noticeably more peaked distribution, with orientations being quite similar in their substructure mass fractions especially for the higher numbers, while in contrast number~20 has a generally flatter slope and a larger variance in the mass fractions of higher number substructures. This can also be seen with regards to the black dotted line representing a value slightly above $2\%$ of the total mass, which for galaxy cluster~5 is only achieved very rarely for substructure number~5 and is common only for substructure~2, while galaxy cluster~20 has this occur even more frequently for substructure numbers as high as~8 and regularly makes this threshold for substructures up to number~5. It is interesting to note the bi-modal behavior in the frequencies of mass fractions which appears for galaxy cluster~20 especially noticeably in the first and third substructure. Here, two distinct peaks are visible, indicating that there exists two classes of orientations for this galaxy cluster: one more commonly with a lower first substructure mass and another more rarely with a nearly~1.5~times as high first mass. Even for the seventh and eighth substructures, there is still a significant spread possible between the masses, unlike for galaxy cluster~5. 

The bottom row of Fig.~\ref{fig:r_5-20} depicts the same for a projection depth of $r_\mathrm{z}=1\cdot r_\mathrm{vir}$ (before and behind the cluster) to test how relevant fore-/ background structures are. Little difference overall is found to the usual projection depth depicted in the top row. The first substructure mass for both clusters rises marginally, with the most noticeable difference being the absence of orientations with a small value for galaxy cluster~5. This implies that there exists in some projections a fore- or background structure which shifts the center-of-mass, thus reducing the first apertures mass in some cases, while the cluster itself is entirely dominated by its brightest cluster galaxy. However, this does not strongly impact the distribution of the higher number substructure masses except for reducing their maxima marginally. This can be seen also for galaxy cluster~20, where the number of projections with an eighth substructure of mass fraction lying above the black line is~3 and~1 for the top and bottom row, respectively. It follows that a deeper projection depth (i.e., less resolved redshift space) does marginally allow for larger peak masses of substructures in the cases where some additional objects are projected into the image, but the overall effect of projection of the galaxy cluster is significantly larger than the impact from these additional structures.

\bibliography{bibliography}

\begin{thebibliography}{65}
\expandafter\ifx\csname natexlab\endcsname\relax\def\natexlab#1{#1}\fi

\bibitem[{{Allen} {et~al.}(2008){Allen}, {Rapetti}, {Schmidt}, {Ebeling},
  {Morris}, \& {Fabian}}]{allen08}
{Allen}, S.~W., {Rapetti}, D.~A., {Schmidt}, R.~W., {et~al.} 2008, \mnras, 383,
  879

\bibitem[{{Bah{\'e}}(2021)}]{bahe21}
{Bah{\'e}}, Y.~M. 2021, \mnras, 505, 1458

\bibitem[{{Bah{\'e}} {et~al.}(2019){Bah{\'e}}, {Schaye}, {Barnes}, {Dalla
  Vecchia}, {Kay}, {Bower}, {Hoekstra}, {McGee}, \& {Theuns}}]{bahe19}
{Bah{\'e}}, Y.~M., {Schaye}, J., {Barnes}, D.~J., {et~al.} 2019, \mnras, 485,
  2287

\bibitem[{{Beck} {et~al.}(2016){Beck}, {Murante}, {Arth}, {Remus}, {Teklu},
  {Donnert}, {Planelles}, {Beck}, {F{\"o}rster}, {Imgrund}, {Dolag}, \&
  {Borgani}}]{beck16}
{Beck}, A.~M., {Murante}, G., {Arth}, A., {et~al.} 2016, \mnras, 455, 2110

\bibitem[{{Bergamini} {et~al.}(2022){Bergamini}, {Acebron}, {Grillo}, {Rosati},
  {Caminha}, {Mercurio}, {Vanzella}, {Angora}, {Brammer}, {Meneghetti}, \&
  {Nonino}}]{bergamini22}
{Bergamini}, P., {Acebron}, A., {Grillo}, C., {et~al.} 2022, arXiv e-prints,
  arXiv:2207.09416

\bibitem[{Besançon {et~al.}(2021)Besançon, Papamarkou, Anthoff, Arslan,
  Byrne, Lin, \& Pearson}]{besancon21}
Besançon, M., Papamarkou, T., Anthoff, D., {et~al.} 2021, Journal of
  Statistical Software, 98, 1

\bibitem[{{Bhattacharyya} {et~al.}(2021){Bhattacharyya}, {Adhikari},
  {Banerjee}, {More}, {Kumar}, {Nadler}, \& {Chatterjee}}]{bhatta21}
{Bhattacharyya}, S., {Adhikari}, S., {Banerjee}, A., {et~al.} 2021, arXiv
  e-prints, arXiv:2106.08292

\bibitem[{{Biffi} {et~al.}(2016){Biffi}, {Borgani}, {Murante}, {Rasia},
  {Planelles}, {Granato}, {Ragone-Figueroa}, {Beck}, {Gaspari}, \&
  {Dolag}}]{biffi16}
{Biffi}, V., {Borgani}, S., {Murante}, G., {et~al.} 2016, \apj, 827, 112

\bibitem[{{Biffi} {et~al.}(2018){Biffi}, {Dolag}, \& {Merloni}}]{biffi18}
{Biffi}, V., {Dolag}, K., \& {Merloni}, A. 2018, \mnras, 481, 2213

\bibitem[{{Bird} \& {Goldberg}(2018)}]{bird18}
{Bird}, J.~P. \& {Goldberg}, D.~M. 2018, \mnras, 476, 1198

\bibitem[{{Boschin} {et~al.}(2006){Boschin}, {Girardi}, {Spolaor}, \&
  {Barrena}}]{boschin06}
{Boschin}, W., {Girardi}, M., {Spolaor}, M., \& {Barrena}, R. 2006, \aap, 449,
  461

\bibitem[{{De Lucia} {et~al.}(2004){De Lucia}, {Kauffmann}, {Springel},
  {White}, {Lanzoni}, {Stoehr}, {Tormen}, \& {Yoshida}}]{delucia04}
{De Lucia}, G., {Kauffmann}, G., {Springel}, V., {et~al.} 2004, \mnras, 348,
  333

\bibitem[{{Dolag} {et~al.}(2009){Dolag}, {Borgani}, {Murante}, \&
  {Springel}}]{dolag09}
{Dolag}, K., {Borgani}, S., {Murante}, G., \& {Springel}, V. 2009, \mnras, 399,
  497

\bibitem[{{Dolag} {et~al.}(2004){Dolag}, {Jubelgas}, {Springel}, {Borgani}, \&
  {Rasia}}]{dolag04}
{Dolag}, K., {Jubelgas}, M., {Springel}, V., {Borgani}, S., \& {Rasia}, E.
  2004, \apjl, 606, L97

\bibitem[{{Dolag} {et~al.}(2017){Dolag}, {Mevius}, \& {Remus}}]{dolag17}
{Dolag}, K., {Mevius}, E., \& {Remus}, R.-S. 2017, Galaxies, 5, 35

\bibitem[{{Dolag} {et~al.}(2005){Dolag}, {Vazza}, {Brunetti}, \&
  {Tormen}}]{dolag05}
{Dolag}, K., {Vazza}, F., {Brunetti}, G., \& {Tormen}, G. 2005, \mnras, 364,
  753

\bibitem[{{Donnert} {et~al.}(2013){Donnert}, {Dolag}, {Brunetti}, \&
  {Cassano}}]{donnert13}
{Donnert}, J., {Dolag}, K., {Brunetti}, G., \& {Cassano}, R. 2013, \mnras, 429,
  3564

\bibitem[{{Eckert} {et~al.}(2015){Eckert}, {Jauzac}, {Shan}, {Kneib}, {Erben},
  {Israel}, {Jullo}, {Klein}, {Massey}, {Richard}, \& {Tchernin}}]{eckert15}
{Eckert}, D., {Jauzac}, M., {Shan}, H., {et~al.} 2015, \nat, 528, 105

\bibitem[{{Eckert} {et~al.}(2016){Eckert}, {Jauzac}, {Vazza}, {Owers}, {Kneib},
  {Tchernin}, {Intema}, \& {Knowles}}]{eckert16}
{Eckert}, D., {Jauzac}, M., {Vazza}, F., {et~al.} 2016, \mnras, 461, 1302

\bibitem[{{Einasto}(1965)}]{einasto65}
{Einasto}, J. 1965, Trudy Astrofizicheskogo Instituta Alma-Ata, 5, 87

\bibitem[{{Fabjan} {et~al.}(2010){Fabjan}, {Borgani}, {Tornatore}, {Saro},
  {Murante}, \& {Dolag}}]{fabjan10}
{Fabjan}, D., {Borgani}, S., {Tornatore}, L., {et~al.} 2010, \mnras, 401, 1670

\bibitem[{{Giovannini} {et~al.}(1999){Giovannini}, {Tordi}, \&
  {Feretti}}]{giovannini99}
{Giovannini}, G., {Tordi}, M., \& {Feretti}, L. 1999, \na, 4, 141

\bibitem[{{Grillo} {et~al.}(2015){Grillo}, {Suyu}, {Rosati}, {Mercurio},
  {Balestra}, {Munari}, {Nonino}, {Caminha}, {Lombardi}, {De Lucia}, {Borgani},
  {Gobat}, {Biviano}, {Girardi}, {Umetsu}, {Coe}, {Koekemoer}, {Postman},
  {Zitrin}, {Halkola}, {Broadhurst}, {Sartoris}, {Presotto}, {Annunziatella},
  {Maier}, {Fritz}, {Vanzella}, \& {Frye}}]{grillo15}
{Grillo}, C., {Suyu}, S.~H., {Rosati}, P., {et~al.} 2015, \apj, 800, 38

\bibitem[{{Harris} {et~al.}(2020){Harris}, {Remus}, {Harris}, \&
  {Babyk}}]{harris20}
{Harris}, W.~E., {Remus}, R.-S., {Harris}, G. L.~H., \& {Babyk}, I.~V. 2020,
  \apj, 905, 28

\bibitem[{{Hirschmann} {et~al.}(2014){Hirschmann}, {Dolag}, {Saro}, {Bachmann},
  {Borgani}, \& {Burkert}}]{hirschmann14}
{Hirschmann}, M., {Dolag}, K., {Saro}, A., {et~al.} 2014, \mnras, 442, 2304

\bibitem[{Hunter(2007)}]{hunter07}
Hunter, J.~D. 2007, Computing in Science \& Engineering, 9, 90

\bibitem[{{Jauzac} {et~al.}(2016){Jauzac}, {Eckert}, {Schwinn}, {Harvey},
  {Baugh}, {Robertson}, {Bose}, {Massey}, {Owers}, {Ebeling}, {Shan}, {Jullo},
  {Kneib}, {Richard}, {Atek}, {Cl{\'e}ment}, {Egami}, {Israel}, {Knowles},
  {Limousin}, {Natarajan}, {Rexroth}, {Taylor}, \& {Tchernin}}]{jauzac16}
{Jauzac}, M., {Eckert}, D., {Schwinn}, J., {et~al.} 2016, \mnras, 463, 3876

\bibitem[{{Jiang} \& {van den Bosch}(2016)}]{jiang16}
{Jiang}, F. \& {van den Bosch}, F.~C. 2016, \mnras, 458, 2848

\bibitem[{{Jiang} \& {van den Bosch}(2017)}]{jiang17}
{Jiang}, F. \& {van den Bosch}, F.~C. 2017, \mnras, 472, 657

\bibitem[{{Kassiola} \& {Kovner}(1993)}]{kassiola93}
{Kassiola}, A. \& {Kovner}, I. 1993, \apj, 417, 450

\bibitem[{{Kempner} \& {David}(2004)}]{kempner04}
{Kempner}, J.~C. \& {David}, L.~P. 2004, \mnras, 349, 385

\bibitem[{{Komatsu} {et~al.}(2011){Komatsu}, {Smith}, {Dunkley}, {Bennett},
  {Gold}, {Hinshaw}, {Jarosik}, {Larson}, {Nolta}, {Page}, {Spergel},
  {Halpern}, {Hill}, {Kogut}, {Limon}, {Meyer}, {Odegard}, {Tucker}, {Weiland},
  {Wollack}, \& {Wright}}]{komatsu11}
{Komatsu}, E., {Smith}, K.~M., {Dunkley}, J., {et~al.} 2011, \apjs, 192, 18

\bibitem[{{Lotz} {et~al.}(2017){Lotz}, {Koekemoer}, {Coe}, {Grogin}, {Capak},
  {Mack}, {Anderson}, {Avila}, {Barker}, {Borncamp}, {Brammer}, {Durbin},
  {Gunning}, {Hilbert}, {Jenkner}, {Khandrika}, {Levay}, {Lucas}, {MacKenty},
  {Ogaz}, {Porterfield}, {Reid}, {Robberto}, {Royle}, {Smith},
  {Storrie-Lombardi}, {Sunnquist}, {Surace}, {Taylor}, {Williams}, {Bullock},
  {Dickinson}, {Finkelstein}, {Natarajan}, {Richard}, {Robertson}, {Tumlinson},
  {Zitrin}, {Flanagan}, {Sembach}, {Soifer}, \& {Mountain}}]{lotz17}
{Lotz}, J.~M., {Koekemoer}, A., {Coe}, D., {et~al.} 2017, \apj, 837, 97

\bibitem[{{Lotz} {et~al.}(2021){Lotz}, {Dolag}, {Remus}, \& {Burkert}}]{lotz21}
{Lotz}, M., {Dolag}, K., {Remus}, R.-S., \& {Burkert}, A. 2021, \mnras, 506,
  4516

\bibitem[{{Lotz} {et~al.}(2019){Lotz}, {Remus}, {Dolag}, {Biviano}, \&
  {Burkert}}]{lotz19}
{Lotz}, M., {Remus}, R.-S., {Dolag}, K., {Biviano}, A., \& {Burkert}, A. 2019,
  \mnras, 488, 5370

\bibitem[{{Mahler} {et~al.}(2018){Mahler}, {Richard}, {Cl{\'e}ment},
  {Lagattuta}, {Schmidt}, {Patr{\'\i}cio}, {Soucail}, {Bacon}, {Pello},
  {Bouwens}, {Maseda}, {Martinez}, {Carollo}, {Inami}, {Leclercq}, \&
  {Wisotzki}}]{mahler18}
{Mahler}, G., {Richard}, J., {Cl{\'e}ment}, B., {et~al.} 2018, \mnras, 473, 663

\bibitem[{{Mao} {et~al.}(2018){Mao}, {Wang}, {Frenk}, {Gao}, {Li}, {Wang},
  {Cao}, \& {Li}}]{mao18}
{Mao}, T.-X., {Wang}, J., {Frenk}, C.~S., {et~al.} 2018, \mnras, 478, L34

\bibitem[{{Martin} {et~al.}(2022){Martin}, {Bazkiaei}, {Iodice}, {Mihos},
  {Montes}, {Benavides}, {Brough}, {Carlin}, {Collins}, {Duc}, {G{\'o}mez},
  {Galaz}, {Hern{\'a}ndez-Toledo}, {Jackson}, {Kaviraj}, {Knapen},
  {Mart{\'\i}nez-Lombilla}, {McGee}, {O'Ryan}, {Prole}, {Rich}, {Rom{\'a}n},
  {Shah}, {Starkenburg}, {Watkins}, {Zaritsky}, {Pichon}, {Armus}, {Bianconi},
  {Buitrago}, {Bus{\'a}}, {Davis}, {Demarco}, {Desmons}, {Garc{\'\i}a},
  {Graham}, {Holwerda}, {Hon}, {Khalid}, {Klehammer}, {Klutse}, {Lazar},
  {Nair}, {Noakes-Kettel}, {Rutkowski}, {Saha}, {Sahu}, {Sola},
  {V{\'a}zquez-Mata}, {Vera-Casanova}, \& {Yoon}}]{martin22}
{Martin}, G., {Bazkiaei}, A.~E., {Iodice}, M. S.~E., {et~al.} 2022, \mnras
  [\eprint[arXiv]{2203.07675}]

\bibitem[{{Meneghetti} {et~al.}(2020){Meneghetti}, {Davoli}, {Bergamini},
  {Rosati}, {Natarajan}, {Giocoli}, {Caminha}, {Metcalf}, {Rasia}, {Borgani},
  {Calura}, {Grillo}, {Mercurio}, \& {Vanzella}}]{meneghetti20}
{Meneghetti}, M., {Davoli}, G., {Bergamini}, P., {et~al.} 2020, Science, 369,
  1347

\bibitem[{{Merten} {et~al.}(2011){Merten}, {Coe}, {Dupke}, {Massey}, {Zitrin},
  {Cypriano}, {Okabe}, {Frye}, {Braglia}, {Jim{\'e}nez-Teja}, {Ben{\'\i}tez},
  {Broadhurst}, {Rhodes}, {Meneghetti}, {Moustakas}, {Sodr{\'e}}, {Krick}, \&
  {Bregman}}]{merten11}
{Merten}, J., {Coe}, D., {Dupke}, R., {et~al.} 2011, \mnras, 417, 333

\bibitem[{{Munari} {et~al.}(2016){Munari}, {Grillo}, {De Lucia}, {Biviano},
  {Annunziatella}, {Borgani}, {Lombardi}, {Mercurio}, \& {Rosati}}]{munari16}
{Munari}, E., {Grillo}, C., {De Lucia}, G., {et~al.} 2016, \apjl, 827, L5

\bibitem[{{Navarro} {et~al.}(1996){Navarro}, {Frenk}, \& {White}}]{navarro96}
{Navarro}, J.~F., {Frenk}, C.~S., \& {White}, S. D.~M. 1996, \apj, 462, 563

\bibitem[{{Neto} {et~al.}(2007){Neto}, {Gao}, {Bett}, {Cole}, {Navarro},
  {Frenk}, {White}, {Springel}, \& {Jenkins}}]{neto07}
{Neto}, A.~F., {Gao}, L., {Bett}, P., {et~al.} 2007, \mnras, 381, 1450

\bibitem[{{Okabe} {et~al.}(2014){Okabe}, {Futamase}, {Kajisawa}, \&
  {Kuroshima}}]{okabe14}
{Okabe}, N., {Futamase}, T., {Kajisawa}, M., \& {Kuroshima}, R. 2014, \apj,
  784, 90

\bibitem[{{Owers} {et~al.}(2011){Owers}, {Randall}, {Nulsen}, {Couch}, {David},
  \& {Kempner}}]{owers11}
{Owers}, M.~S., {Randall}, S.~W., {Nulsen}, P. E.~J., {et~al.} 2011, \apj, 728,
  27

\bibitem[{{Power} {et~al.}(2003){Power}, {Navarro}, {Jenkins}, {Frenk},
  {White}, {Springel}, {Stadel}, \& {Quinn}}]{power03}
{Power}, C., {Navarro}, J.~F., {Jenkins}, A., {et~al.} 2003, \mnras, 338, 14

\bibitem[{{Ragagnin} {et~al.}(2017){Ragagnin}, {Dolag}, {Biffi}, {Cadolle Bel},
  {Hammer}, {Krukau}, {Petkova}, \& {Steinborn}}]{ragagnin17}
{Ragagnin}, A., {Dolag}, K., {Biffi}, V., {et~al.} 2017, Astronomy and
  Computing, 20, 52

\bibitem[{{Ragagnin} {et~al.}(2019){Ragagnin}, {Dolag}, {Moscardini},
  {Biviano}, \& {D'Onofrio}}]{ragagnin19}
{Ragagnin}, A., {Dolag}, K., {Moscardini}, L., {Biviano}, A., \& {D'Onofrio},
  M. 2019, \mnras, 486, 4001

\bibitem[{{Ragagnin} {et~al.}(2021){Ragagnin}, {Fumagalli}, {Castro}, {Dolag},
  {Saro}, {Costanzi}, \& {Bocquet}}]{ragagnin21}
{Ragagnin}, A., {Fumagalli}, A., {Castro}, T., {et~al.} 2021, arXiv e-prints,
  arXiv:2110.05498

\bibitem[{{Ragagnin} {et~al.}(2022){Ragagnin}, {Meneghetti}, {Bassini},
  {Ragone-Figueroa}, {Granato}, {Despali}, {Giocoli}, {Granata}, {Moscardini},
  {Bergamini}, {Rasia}, {Valentini}, {Borgani}, {Calura}, {Dolag}, {Grillo},
  {Mercurio}, {Murante}, {Natarajan}, {Rosati}, {Taffoni}, {Tornatore}, \&
  {Tortorelli}}]{ragagnin22}
{Ragagnin}, A., {Meneghetti}, M., {Bassini}, L., {et~al.} 2022, arXiv e-prints,
  arXiv:2204.09067

\bibitem[{{Rajpurohit} {et~al.}(2021){Rajpurohit}, {Vazza}, {van Weeren},
  {Hoeft}, {Brienza}, {Bonnassieux}, {Riseley}, {Brunetti}, {Bonafede},
  {Br{\"u}ggen}, {Formann}, {Rajpurohit}, {R{\"o}ttgering}, {Drabent},
  {Dom{\'\i}nguez-Fern{\'a}ndez}, {Wittor}, \& {Andrade-Santos}}]{rajpurohit21}
{Rajpurohit}, K., {Vazza}, F., {van Weeren}, R.~J., {et~al.} 2021, \aap, 654,
  A41

\bibitem[{{Remus} {et~al.}(2022){Remus}, {Dolag}, \& {Dannerbauer}}]{remus22_2}
{Remus}, R.-S., {Dolag}, K., \& {Dannerbauer}, H. 2022, arXiv e-prints,
  arXiv:2208.01053

\bibitem[{{Remus} {et~al.}(2017){Remus}, {Dolag}, \& {Hoffmann}}]{remus17_2}
{Remus}, R.-S., {Dolag}, K., \& {Hoffmann}, T. 2017, Galaxies, 5, 49

\bibitem[{{Remus} \& {Forbes}(2022)}]{remus22}
{Remus}, R.-S. \& {Forbes}, D.~A. 2022, \apj, 935, 37

\bibitem[{{Retana-Montenegro} {et~al.}(2012){Retana-Montenegro}, {van Hese},
  {Gentile}, {Baes}, \& {Frutos-Alfaro}}]{retana12}
{Retana-Montenegro}, E., {van Hese}, E., {Gentile}, G., {Baes}, M., \&
  {Frutos-Alfaro}, F. 2012, \aap, 540, A70

\bibitem[{{Schulze} {et~al.}(2018){Schulze}, {Remus}, {Dolag}, {Burkert},
  {Emsellem}, \& {van de Ven}}]{schulze18}
{Schulze}, F., {Remus}, R.-S., {Dolag}, K., {et~al.} 2018, \mnras, 480, 4636

\bibitem[{{Schwinn} {et~al.}(2018){Schwinn}, {Baugh}, {Jauzac}, {Bartelmann},
  \& {Eckert}}]{schwinn18}
{Schwinn}, J., {Baugh}, C.~M., {Jauzac}, M., {Bartelmann}, M., \& {Eckert}, D.
  2018, \mnras, 481, 4300

\bibitem[{{Schwinn} {et~al.}(2017){Schwinn}, {Jauzac}, {Baugh}, {Bartelmann},
  {Eckert}, {Harvey}, {Natarajan}, \& {Massey}}]{schwinn17}
{Schwinn}, J., {Jauzac}, M., {Baugh}, C.~M., {et~al.} 2017, \mnras, 467, 2913

\bibitem[{{Springel}(2005)}]{springel05}
{Springel}, V. 2005, \mnras, 364, 1105

\bibitem[{{Takada} \& {Jain}(2003)}]{takada03}
{Takada}, M. \& {Jain}, B. 2003, \mnras, 340, 580

\bibitem[{{Teklu} {et~al.}(2015){Teklu}, {Remus}, {Dolag}, {Beck}, {Burkert},
  {Schmidt}, {Schulze}, \& {Steinborn}}]{teklu15}
{Teklu}, A.~F., {Remus}, R.-S., {Dolag}, K., {et~al.} 2015, \apj, 812, 29

\bibitem[{{Tornatore} {et~al.}(2007){Tornatore}, {Borgani}, {Dolag}, \&
  {Matteucci}}]{tornatore07}
{Tornatore}, L., {Borgani}, S., {Dolag}, K., \& {Matteucci}, F. 2007, \mnras,
  382, 1050

\bibitem[{{Tornatore} {et~al.}(2004){Tornatore}, {Borgani}, {Matteucci},
  {Recchi}, \& {Tozzi}}]{tornatore04}
{Tornatore}, L., {Borgani}, S., {Matteucci}, F., {Recchi}, S., \& {Tozzi}, P.
  2004, \mnras, 349, L19

\bibitem[{{Treu} {et~al.}(2022){Treu}, {Roberts-Borsani}, {Bradac}, {Brammer},
  {Fontana}, {Henry}, {Mason}, {Morishita}, {Pentericci}, {Wang}, {Acebron},
  {Bagley}, {Bergamini}, {Belfiori}, {Bonchi}, {Boyett}, {Boutsia},
  {Calabr{\'o}}, {Caminha}, {Castellano}, {Dressler}, {Glazebrook}, {Grillo},
  {Jacobs}, {Jones}, {Kelly}, {Leethochawalit}, {Malkan}, {Marchesini},
  {Mascia}, {Mercurio}, {Merlin}, {Nanayakkara}, {Nonino}, {Paris},
  {Poggianti}, {Rosati}, {Santini}, {Scarlata}, {Shipley}, {Strait}, {Trenti},
  {Tubthong}, {Vanzella}, {Vulcani}, \& {Yang}}]{treu22}
{Treu}, T., {Roberts-Borsani}, G., {Bradac}, M., {et~al.} 2022, \apj, 935, 110

\bibitem[{{Wiersma} {et~al.}(2009){Wiersma}, {Schaye}, \& {Smith}}]{wiersma09}
{Wiersma}, R.~P.~C., {Schaye}, J., \& {Smith}, B.~D. 2009, \mnras, 393, 99

\end{thebibliography}
\bibliographystyle{aa}

\end{document}